\documentstyle[12pt]{report}
\input epsf.sty

\begin{document}

\title{{\Large {\bf THERMAL DESCRIPTION OF PARTICLE PRODUCTION IN
ULTRA-RELATIVISTIC HEAVY-ION COLLISIONS} } }
\vspace{1cm}
\author{by 
\\
\\ 
Mariusz Michalec 
\\
\\
DISSERTATION \\
submitted in partial fulfillment of the requirements \\
for the degree \\
DOCTOR OF PHILOSOPHY IN PHYSICS
\\
\\
THE H. NIEWODNICZA\'NSKI
\\
INSTITUTE OF NUCLEAR PHYSICS, KRAK\'OW
\\
THEORY DEPARTMENT
\\
SUPERVISOR: Assoc. Prof. Wojciech Florkowski}

\maketitle

\newpage
\thispagestyle{empty}
$\mbox{}$

\newpage
\thispagestyle{empty}

\vspace{4cm}
\centerline{\bf ABSTRACT}
\vspace{2cm}

The grand-canonical version of the thermal model is used to analyze
the ratios of particle abundances measured in ultra-relativistic
heavy-ion collisions. Exactly the same model is applied to study the
heavy-ion reactions at BNL AGS, CERN SPS, and BNL RHIC. A very good
description is achieved for Pb + Pb collisions at CERN SPS and for Au
+ Au collisions at BNL RHIC. In these two cases the value of the
temperature characterizing the chemical freeze-out is practically the
same: we find $T_{chem}=168 \pm 3$ MeV at SPS, and $T_{chem}=165 \pm
7$ MeV at RHIC. On the other hand, the particle ratios measured in the
collisions of lighter nuclei are described only in the qualitative
way. We discuss also the effect of the possible in-medium
modifications of hadron masses and widths on the thermal fits. For Pb
+ Pb collisions at CERN SPS and Au + Au collisions at BNL RHIC, we
find that the $\chi^2$ fits favor slightly a moderate, $\sim$ 20\%, decrease of
the masses. In this case, the fits with the modified masses yield modified
values of the optimal temperature and the baryon chemical
potential. In-medium modifications of the widths have little effect on
the fits, unless they are increased by a factor larger than 2.  We
study in detail the thermodynamic conditions characterizing the
chemical freeze-out. In particular we find that the average baryon
energy at freeze-out is 1.6 GeV, and the average meson energy is 0.9
GeV. This difference reflects a different behavior of the mass spectra
of baryons and mesons. Similarities and differences between our
calculations and other studies are thoroughly discussed.

\newpage
\thispagestyle{empty}
$\mbox{}$

\setcounter{page}{4}

\tableofcontents

\newsavebox{\tabRHIC}

\chapter{Introduction}

\section{Ultra-Relativistic Heavy-Ion Collisions}

The field of the ultra-relativistic heavy-ion collisions connects the
nuclear physics with the elementary particle physics. In the
high-energy particle physics the interactions are derived from first
principles ({\it local gauge theories}) and one deals with single
particles ({\it leptons, hadrons}, or {\it quarks and gluons}). On the
other hand, in the traditional nuclear physics the strong interaction
is described by {\it effective models}, and the matter consists of
extended complicated systems ({\it nuclei}). A unifying aspect of the
high-energy nuclear collisions is an attempt to analyze the properties
of dense hadronic matter in terms of elementary interactions. The
fundamental theory of strong interactions, {\it Quantum
Chromodynamics} (QCD), predicts that at high energy density, hadronic
matter will turn into a plasma of deconfined quarks and gluons
(QGP). The search for such a phase transition is the main motivation
for the continuous experimental and theoretical efforts in the field
%\cite{QM90,QM91,QM93,QM95,QM96,QM97,QM99,QM01}
[1-8].  One of the goals of
the experimental program is to recreate (on a microscopic scale) the
physical conditions that are thought to have existed in the early
universe. Such astrophysical aspect of the ultra-relativistic
heavy-ion collisions indicates once again on the interdisciplinary
character of this new branch of physics.

The experimental studies of the high-energy nuclear collisions started
in 1986 at the Brookhaven National Laboratory (BNL) and at the
European Organization for Nuclear Research (CERN). The Alternating
Gradient Synchrotron (AGS) at BNL accelerated $^{28}$Si beams (at 15
GeV per nucleon) and later $^{197}$Au beams (at 11 GeV per
nucleon). At CERN the Super Proton Synchrotron (SPS) delivered
$^{16}$O and $^{32}$S beams (at 60 and later at 200 GeV per nucleon),
which were followed in 1995 by $^{208}$Pb beams (at 158 GeV per
nucleon). The new era in the field started in 2000, when the
Relativistic Heavy Ion Collider (RHIC) started operation at BNL.  In
the first run of RHIC, Au on Au reactions were studied at the
center-of-mass energy $\sqrt{s}=$ 56 A GeV and $\sqrt{s}=$ 130 A GeV.
In 2001, during the second run the full collision energy, $\sqrt{s}=$
200 A GeV, and the full luminosity were achieved.

Over the last 15 years, the large amount of data has been accumulated.
In particular, the data indicate that the hadronic matter at {\it
freeze-out} (i.e., at the moment when the hadron interactions cease
and the particles freely stream from the collision point to the
detectors) is well described by the equilibrium distributions
%\cite{PBM-AGS,PBM-SPS,CleyEllSa,RafLetTou,PBMHS,CleyRed,YenGor,BecCleyKSR1,
%BecCleyKSR2,Gaz,nxu,pbmrhic,FlorBronMich-TherAnal,sh,ptspec}
[9-23]. In addition, as
one moves up from SPS to RHIC energy, approximately the same
temperature and a significantly smaller baryonic chemical potential is
observed in the central rapidity region.  This fact constrains
different possible models of the particle production.  Actually, there
is no single space-time model describing the whole collision process,
since different degrees of freedom are important at different stages
of a collision. As a consequence, each stage requires a different
treatment: The initial moments are described with the help of partonic
degrees of freedom (quarks and gluons). The intermediate stage is
described typically in the framework of the relativistic hydrodynamics
(in this case the degrees of freedom are collectively included in the
equation of state).  Finally, the last stage is described in purely
hadronic models.

\begin{table}[h]
\begin{center}
\begin{tabular}{|c|c|c|c|c|c|}
\hline
& & & & & \\
place & projectile & target(s) & $E_{lab}/A\,\,$[GeV] & 
$y_{lab}$ & $\sqrt{s}/A$\thinspace \thinspace \lbrack GeV] \\
& & & & & \\ \hline\hline
& & & & & \\
\multicolumn{1}{|l|}{BNL \ AGS} & Si & Au, Pb & 14.6 & 3.4 & 5.4 \\ 
& & & & & \\
\hline
& & & & & \\
\multicolumn{1}{|l|}{CERN \ SPS} & S & Au, W, Pb & 200 & 6.1 & 19.4 \\ 
& & & & & \\
\hline
& & & & & \\
\multicolumn{1}{|l|}{CERN \ SPS} & Pb & Pb & 158 & 5.8 & 17.3 \\ 
& & & & & \\
\hline
& & & & & \\
\multicolumn{1}{|l|}{BNL \ RHIC} & Au & Au & $\left( 
\begin{tabular}{c}
9000 \\ 
21300
\end{tabular}
\right) $ & $\left( 
\begin{tabular}{c}
9.9 \\ 
10.7
\end{tabular}
\right) $ & 
\begin{tabular}{l}
130 \\ 
200
\end{tabular}
\\ 
& & & & & \\
\hline
\end{tabular}
\end{center}
\caption{{\small Heavy-ion reactions studied in this paper. }}
\end{table}

\section{Aim of this Work}

In this work we discuss the properties of matter at freeze-out, thus
we are not interested in the microscopic mechanism of particle
production, such as decaying color strings \cite{tubes} or a parton
cascade \cite{cascade}. We assume, however, that the real microscopic
mechanism leads finally to creation of a {\it locally equilibrated
hadron gas}. We note that in our approach one cannot draw any direct
conclusions concerning formation of the quark-gluon plasma at the
earlier stages of a collision. Such information is lost in the thermal
equilibrium, which has no memory about earlier times.

Our main purpose is to perform a thermal analysis of the particle
yields. We are going to check whether {\it the measured particle
yields may be explained in a model which assumes full thermal and
chemical equilibrium of the hadronic matter at freeze-out}. In the
last years such thermal approach has been used by several groups to
study different types of reactions. The results of such investigations
show that the thermal description of the particle production is quite
successful. We have to have in mind, however, that different groups
use different implementations of the model, and these implementations
vary for each particular reaction. In this situation, the conclusions
drawn from the vast applications of the thermal models may be not
completely consistent. In this paper, to minimize such effects, one
formulation of the model is used to describe most of the available
data. In this way, exactly the same thermodynamic features of the
particle production are used to characterize different
collisions. This allows us to observe similarities and differences in
the thermodynamic behavior of various colliding systems.

The second and independent aim of this work is to include the possible
in-medium modifications of hadron properties into the thermal
approach. In the scenario with the two different freeze-outs, i.e.,
with the chemical freeze-out preceding the thermal freeze-out, the
study of the particle ratios reveals the information about the
chemical freeze-out: the concept of the chemical freeze-out refers to
the point at which inelastic collisions cease and all particle ratios
are frozen, whereas the concept of the thermal freeze-out refers to
the stage when all elastic collisions cease.  If the hadronic matter
at the chemical freeze-out is very hot and dense, we may take into
account the mass and width modifications of hadrons. Such
modifications are predicted by the effective theories of QCD
%\cite{Hirsch95,Hirsch00,hatsuda,klingl}
[26-29], and also by the QCD sum rules
in medium \cite{HatsudaLee}. For instance, according to Brown and Rho
\cite{BR}, the masses of hadrons decrease at higher densities. The
change of the mass leads to a change of the hadron densities, which
should be reflected in the measured relative particle yields. The
study of such effects is the second important issue discussed in the
present paper.

In our study of the particle multiplicities we use our own
implementation of the thermal model, which has been constructed as a
code in the {\small MATHEMATICA} language. The use of the symbolic
manipulations allowed by the {\small MATHEMATICA }turned out to be
very convenient in the treatment of the hadronic decays. Creation of
this code was the main technical task connected with the present
investigations.  A part of the original results discussed in this
paper has been published before in the two articles:
\vspace{-0.15cm}
\begin{itemize}
\item[I.] M. Michalec, W. Florkowski, and W. Broniowski:
{\it Scaling of hadron masses and widths in thermal models
for ultra-relativistic heavy-ion collisions}, Phys. Lett.
{\bf B520} (2001) 213; nucl-th/0103029.
\item[II.] W. Florkowski, W. Broniowski, and M. Michalec:
{\it Thermal analysis of particle ratios and $p_\perp$ spectra
at RHIC}, nucl-th/0106009.
\end{itemize}

\newpage
\thispagestyle{empty}
$\mbox{}$

\chapter{Thermal Model of Particle Production}

\section{Historical Perspective}

The use of statistical concepts to describe particle production in
hadronic collisions has a long history. In the early fifties Fermi
\cite{Fermi} assumed that when two relativistic nucleons collide, the
energy available in their center-of-mass system is released in a very
small volume (due to the Lorentz contraction).  Subsequently, such a
dense system decays into one of many accessible multiparticle states.
The decay probabilities were calculated by Fermi according to the
standard rules of statistical mechanics. In the Landau's hydrodynamic
model the picture of the instantaneous break-up was modified by the
inclusion of the expansion stage. The volume expansion leads to
cooling and lowering of the freeze-out temperature, and causes that
most final hadrons are the light ones.

In the Landau hydrodynamic model \cite{Landau} the initial conditions
are specified at a given time, when the matter is highly compressed
and at rest.  This description does not include one aspect of the
particle production at high energies, namely, the fact that fast
particles are produced later and farther away from the collision
center than slow particles.  This feature of particle production has
been included in the Bjorken \cite{Bjorken,Baym} hydrodynamic model
which imposes boost-invariant initial conditions.

Thermal and/or statistical interpretation of the particle production
became a common approach for the ultra-relativistic heavy-ion collisions
%\cite{PBM-AGS,CleyEllSa,PBM-SPS,RafLetTou,PBMHS,CleyRed,YenGor,BecCleyKSR1,
%BecCleyKSR2,Gaz,pbmrhic,FlorBronMich-TherAnal,sh,ptspec}
[9-23].  In this case large
numbers ({\it multiplicities}) of hadrons are created, and the
statistical methods seem to be appropriate. Of course, the produced
matter may be formed in a state which is far away from the local
equilibrium, so the use of the simple equilibrium concepts must be
grounded in more involved studies using, e.g., the kinetic theory of
hadronic matter.

\section{Thermal and Chemical Freeze-Out}

In the simplest version of the thermal model one assumes that the
hadronic matter created in nuclear collisions forms an ideal gas. A
volume expansion of the gas, caused by high internal pressure, leads
to {\it decoupling} or {\it freeze-out} of the hadrons. This process
takes place when the mean free path of hadrons becomes compatible with
the macroscopic size of the whole hadronic system. After the
freeze-out, the particles move freely and their properties may be
measured in the detectors. If the freeze-out process is fast, the
momentum distributions of the particles do not change substantially,
and the measured spectra have thermal shapes. Thus, by studying the
hadron momentum distributions we may check the validity of the concept
of thermalization. Moreover, in the thermal model the abundances of
different hadron species are fixed by the values of a few
thermodynamic parameters. Consequences of this fact can be verified by
the measurements of the {\it relative particle yields}.

More realistically, the hadronic system at the freeze-out can be
viewed as a collection of subsystems. Each of such subsystems can be
characterized by the individual (or {\it local}) values of the
thermodynamic parameters, and by the individual value of the
collective velocity. The case described above corresponds to the
situation when all local thermodynamic parameters are the same, and
the collective velocities are small. In this approximation the
measurements of the momentum distributions and of the relative yields
should reveal the same value of the temperature and the same values of
the chemical potentials. On the other hand, if the collective
velocities are not negligible, the momentum spectra of hadrons are
modified by a superposition of ''redshift'' and ''blueshift''
effects. The spectra do not reveal the {\it true local} freeze-out
conditions anymore. It can been shown, however, that the measurement
of the relative particle yields is not affected by the collective
velocities, provided the local thermodynamic parameters are the same
\cite{Heinz,CleyOesRed}. This crucial observation makes grounds for a
large interest in the studies of the particle ratios. In this paper we
follow this trend and perform thermal analysis of the particle ratios
measured at BNL AGS, CERN\ SPS, and BNL\ RHIC.

Since the thermal fits to the particle ratios give quite large values
of the optimal temperature, one finds typically $T\sim $ 170 MeV, a
concept of the two different freeze-outs has been introduced: at first
the {\it chemical freeze-out} takes place, and only later the{\it \
thermal} (or {\it kinetic}) freeze-out happens. At the chemical
freeze-out the chemical contents of the hadronic system is
established. Later, only elastic processes are possible (dominate),
leading to further expansion of the system and cooling.  Finally, when
the system becomes sufficiently diluted, the true thermal freeze-out
takes place (as described above). The concept of the two freeze-outs
helps to understand the large difference between the temperature
inferred from the investigation of the particle ratios, and the
temperature inferred from the study of the momentum distributions. At
SPS\ energies, the temperature of the thermal freeze-out \ is around
$T\sim $130 MeV. A detailed study of the freeze-out mechanism can be
performed only in the microscopic framework, such as the relativistic
kinetic theory. {\it In our approach, we simply adopt the definition
of the chemical freeze-out and check if the particle multiplicities
measured in the ultra-relativistic heavy-ion collisions are consistent
with this idea.}

It should be noticed, however, that there exist models of hadron
production in ultra-relativistic heavy-ion collisions, which assume
that the chemical freeze-out coincides with the thermal freeze-out. An
example of such a model is the {\it sudden hadronization} scenario of
Ref. \cite{sh}. Another model with a single freeze-out is defined in
Ref. \cite{ptspec}.  We want to emphasize that in this case (i.e., in
the situations when a single freeze-out is considered) our
calculations are also useful for the determination of the
thermodynamic properties of hadronic matter. For example, the results
of our analysis of the particle ratios at RHIC
\cite{FlorBronMich-TherAnal} were combined with the hydrodynamic
expansion in order to calculate the transverse-momentum spectra of
hadrons \cite{ptspec}.  This approach yields very successful fits and
shows that the thermal analysis of the ratios may be the first step in
more complex investigations of other observables such as elliptic flow
or HBT radii.

\begin{figure}[h]
\epsfysize=8cm
\par
\begin{center}
\mbox{\epsfbox{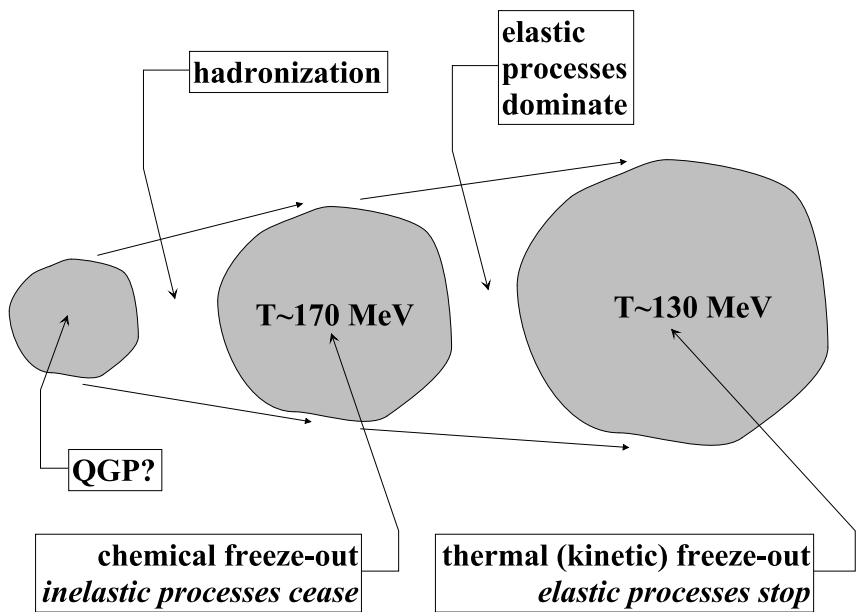}}
\end{center}
\caption{{\small In many scenarios of the evolution of the  hadronic matter
in ultra-relativistic heavy-ion collisions, the chemical freeze-out
precedes the thermal (kinetic) freeze-out. At the chemical freeze-out
the inelastic processes stop and the chemical content of the fireball
is established. Later on, only elastic processes are present. They lead
to further expansion and cooling of the system. At the thermal freeze-out
the elastic interactions cease, and the particles stream freely to
detectors. A high temperature of the chemical freeze-out suggests that it
happens just after hadronization of the quark-gluon plasma possibly created
in the collisions. }
}
\label{fout}
\end{figure}

\section{Basics of Thermal Analysis}

\subsection{Static Fireball}

In this Section we give the basic formulas used in the thermal analysis of
the particle ratios. We start with the presentation of the simplest approach
and assume that the (chemical) freeze-out takes place in a static volume. In
this case the multiplicities of hadrons of species $i$ are obtained from the
ideal-gas expression 
\begin{equation}
N_{i} = V n_i = Vg_{i}\int d^{3}p\ f_{i}\left( p\right).  
\label{Ni}
\end{equation}
Here $V$ is the volume of the hadronic system at the freeze-out, $%
g_{i}=2J_{i}+1$ is the spin degeneracy factor of the $i$th hadron, and $%
f_{i}(p)$ is the momentum distribution function. In the thermodynamic
equilibrium the distribution functions $f_{i}(p)$ have a form ($\hbar=1$)
\begin{equation}
f_{i}\left( p\right) =\frac{1}{\left( 2\pi \right) ^{3}}\left[ \exp \left( 
\frac{E_{i}(p)-\mu _{i}}{T}\right) +\epsilon \right] ^{-1},
\label{fip}
\end{equation}
where 
\begin{equation}
E_{i}(p)=\sqrt{p^{2}+m_{i}^{2}}  
\label{Eip}
\end{equation}
is the energy, $T$ is the temperature, and $\mu _{i}$ is the chemical
potential. The quantity $\epsilon $ equals +1 for fermions (in this
case (\ref{fip}) becomes the Fermi-Dirac distribution) and -1 for
bosons (in this case (\ref{fip}) becomes the Bose-Einstein
distribution).  The limit $\epsilon \longrightarrow 0$ corresponds to
the classical (Boltzmann) statistics. The chemical potential $\mu
_{i}$ in Eq. (\ref{fip}) is a linear combination of the baryon,
strange and isospin chemical potential

\begin{equation}
\mu _{i}=\mu ^{B}B_{i}+\mu ^{S}S_{i}+\mu ^{I}I_{i}.  \label{miu}
\end{equation}
Here $B_{i},S_{i},$ and $I_{i}$ are the baryon number, strangeness,
and the third component of isospin of the $i$th hadron, respectively.

Introduction of the chemical potentials $\mu ^{B},\mu ^{S}$ and $\mu
^{I}$, allows us to fulfill the appropriate conservation laws. We
assume that the strangeness of the system is zero
\begin{equation}
\sum_{i}S_{i}N_{i}=0,  
\label{s0}
\end{equation}
and the ratio of the electric charge to the baryon number in the
hadronic fireball is the same as in the colliding nuclei
\begin{equation}
\frac{\sum_{i}Q_{i}N_{i}}{\sum_{i}B_{i}N_{i}}=\frac{Z}{A}.  
\label{zovera}
\end{equation}
Here $Q_{i}$ is the electric charge of the $i$th hadron. According to the
Gell-Mann -- Nishijima formula we have 
\begin{equation}
Q_{i}=I_{i}+\frac{\left( B_{i}+S_{i}\right) }{2}. 
\label{gmn}
\end{equation}
In the practical calculations, Eqs. (\ref{s0}) and (\ref{zovera}) are
used to fix the values of the chemical potentials $\mu ^{S}$ and $\mu
^{I}$. Thus, we are left with only two independent parameters: the
temperature and the baryon chemical potential. We note that the volume
cancels in the conditions (\ref{s0})\ and (\ref{zovera}), and also in
the particle ratios introduced below. The values of $Z$ and $A$ (for
the nuclei discussed in our paper) are given in the Table \ref{ZA}.

\begin{table}[h]
\begin{center}
\begin{tabular}{|l|l|l|l|l|l|}
\hline
& Si & S & W & Au & Pb \\ \hline\hline
$Z$ & 14 & 16 & 74 & 79 & 82 \\ \hline
$A$ & 29 & 32 & 184 & 197 & 207 \\ \hline
$Z/A$ & 0.48 & 0.50 & 0.40 & 0.40 & 0.40 \\ \hline
\end{tabular}
\end{center}
\caption{{\small The values of the atomic numbers, $Z$, 
and the mass numbers, $A$, 
for the nuclei discussed in the present paper.}}
\label{ZA}
\end{table}

\subsection{Expanding Fireball}

\begin{figure}[t]
\epsfysize=8cm
\par
\begin{center}
\mbox{\epsfbox{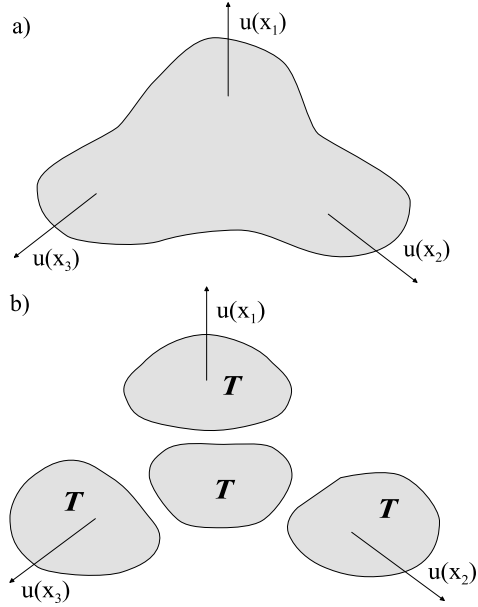}}
\end{center}
\caption{{\small Illustration of the insensitivity of the particle
ratios to hydrodynamic flow. An expanding system (a) can be viewed as
a collection of subsystems (b) which move with collective velocities
$u(x)$. If the thermodynamic potentials are the same in each
subsystem, the particles are emitted in identical proportions. In this
case the measurement of the yields in the full phase-space reveals the
local thermodynamic conditions.}}
\label{fball2}
\end{figure}

In a more general case, when the expansion of the system at freeze-out
cannot be neglected, one may use the Cooper-Frye formula
\cite{CooperFrye,CooperFryeSchon} to calculate the total yield of
particles of species $i$, namely
\begin{equation}
N_{i}=\int \frac{d^{3}p}{E_{i}(p)}\int d\Sigma _{\mu }\,\,p^{\mu }\,\,g_{i}\
f_{i}\left( \sqrt{\left( p\cdot u\right) ^{2}-m_{i}^{2}}\right).
\label{CF}
\end{equation}
Here $\Sigma _{\mu }$ describes the element of the freeze-out
hyper-surface, and $u^\mu$ is the local hydrodynamic four-velocity. In
the local rest frame: $u^\mu = (1,0,0,0)$ and $\sqrt{\left( p\cdot
u\right) ^{2}-m_{i}^{2}}=p$.  Both $\Sigma _{\mu }$ and $u^\mu$ depend
on the space-time position $x$. In general, also $T$ and $\mu_i$ may
be defined locally.

Eq. (\ref{CF}) can be rewritten in terms of
the number current \cite{Heinz}
\begin{equation}
N_{i}=\int d\Sigma _{\mu }(x) \,j_{i}^{\mu }(x),
\end{equation}
where
\begin{eqnarray}
j_{i}^{\mu }\left( x\right)  &=&2 \int d^{4}p\,\theta \left( p^{0}\right)
\delta \left( p^{2}-m_{i}^{2}\right) p^{\mu }\,\,g_{i}\ \,f_{i}\left( \sqrt{%
\left( p\cdot u\right) ^{2}-m_{i}^{2}}\right)  \nonumber \\
&=&2 \int d^{4}p\,\theta \left( p^{0}\right) \delta \left(
p^{2}-m_{i}^{2}\right) p^{\mu }\,\,
g_{i}
\left[\exp\left({p\cdot u-\mu_{i} \over T} \right) + \epsilon \right]^{-1} .
\label{jmu}
\end{eqnarray}
Eq. (\ref{jmu}) is written in a manifestly Lorentz-covariant way. The
step function $\theta(x)$ is defined by the conditions
\begin{equation}
\theta(x) = 1 \,\,\, \hbox{for} \,\,\, x \ge 0, \,\,\,
\hbox{and} \,\,\, \theta(x) = 0 \,\,\, \hbox{for} \,\,\, x < 0.
\end{equation}
In local thermal equilibrium the number current is proportional to the
four-velocity,
\begin{equation}
j_{i}^{\mu }\left( x\right) =\rho _{i}\left( x\right) \,u^{\mu }\left(
x\right),
\end{equation}
and
\begin{eqnarray}
\rho _{i}\left( x\right)& = & u_{\mu }\left( x\right) \,\,j_{i}^{\mu }\left(
x\right) \nonumber  \\
&=&2 \int d^{4}p\,\theta \left( p^{0}\right) \delta \left(
p^{2}-m_{i}^{2}\right) p^{\mu }\cdot u_{\mu }\,\,g_{i}\ \,f_{i}\left( \sqrt{%
\left( p\cdot u\right) ^{2}-m_{i}^{2}}\right) \nonumber  \\
&=&g_{i}\int d^{3}p\ f_{i}\left( p\right) =n_{i}\left( T \left(
x\right) ,\mu^{i}\left( x\right) \right).
\end{eqnarray}
Here $n_i(x)$ denotes the equilibrium particle density at the
temperature $T(x)$ and the chemical potential $\mu_{i}(x)$.  The total
particle yield of species $i$ is therefore
\begin{equation}
N_{i} =  \int d\Sigma _{\mu }\,
n_{i}\left( T\left( x\right) ,\mu_{i}\left( x\right)
\right) \,u^{\mu }\left( x\right).
\label{locni}
\end{equation}
We observe that the particle ratios do not depend on a particular shape
of the freeze-out surface as long as the local thermodynamic parameters
are independent of $x$. In this case we have
\begin{equation}
{N_i \over N_j} = 
{ n_{i} (T,\mu_{i})\int d\Sigma _{\mu }\,u^{\mu }\left( x\right)   
\over 
n_{j} (T,\mu_{j}) \int  d\Sigma _{\mu }\,u^{\mu }\left( x\right) } =  
{ n_{i} (T,\mu_{i})  \over 
n_{j} (T,\mu_{j})},
\end{equation}
so the ratios are the same as those in a static fireball.

The case considered above should be confronted with the analysis
of the particle spectra. The latter are strongly influenced by the
hydrodynamic flow, since the flow changes the individual velocities of
the particles. As a consequence, the measurement of so-called inverse
slopes of the $p_\perp$ spectra does not bring the direct information
about the real temperature of the emitting hadronic source.  For
example, the pieces of hadronic matter moving towards the observer
(large transverse flow at zero rapidity) seem to be hotter than those
being at rest (small transverse flow). This is a kind of the
blueshift effect, which effectively raises the measured inverse slope.

Strictly speaking, the arguments presented in this Section show that
the particle ratios are insensitive to the hydrodynamic flow if {\it
all} rapidities and transverse momenta of the particles are measured
(in Eqs. (\ref{CF}) and (\ref{jmu}) one integrates over all momenta
{\bf p}). From the experimental point of view, this requires the full
$4 \pi$ acceptance of the detectors. Nevertheless, this condition
may be relaxed for boost-invariant systems. In this case $dN_i/dy$
is independent of $y$ and we have \cite{ptspec}
\begin{equation}
{dN_i/dy \over dN_j/dy} = {\int dy \, dN_i/dy \over \int dy \, dN_j/dy}
= {N_i \over N_j}.
\label{niynjy}
\end{equation}
Here
\begin{equation}
{dN_i \over dy} = \int d^2p_\perp \, {dN_i \over d^2p_\perp \, dy}.
\label{niy}
\end{equation}
Thus, for the boost-invariant systems it is enough to measure all particles
at midrapidity, i.e., for $y \approx 0$. The measurements at midrapidity are
also sufficient for the systems which are approximately boost-invariant
\cite{ptspec}. This property will be used in our analysis of the particle
ratios measured in Au + Au collisions at RHIC.

\subsection{Decays of Resonances}
\label{dor}

The hadronic fireball consists of stable hadrons (with respect to
strong interactions) and all hadronic resonances.  In our analysis we
use the newest edition of the review of particle physics \cite{PDG}.
Practically, we include all light-flavor hadrons, i.e., hadrons
containing $u,d$ and $s$ quarks. A few hadrons are not taken into
account, since their properties are not known sufficiently well. All
together we include 164 baryonic states and 110 mesonic states
(treating separately different isospin states). When calculating the
relative yields of the measured hadrons we should include all decay
channels. This may be represented schematically by the expression
\begin{equation}
R=\frac{N_{i}+\sum_{k}b\left( k\rightarrow i\right) N_{k}+\sum_{kl}b\left(
k\rightarrow l\right) b\left( l\rightarrow i\right) N_{k}+...}{%
N_{j}+\sum_{k}b\left( k\rightarrow j\right) N_{k}+\sum_{kl}b\left(
k\rightarrow l\right) b\left( l\rightarrow j\right) N_{k}+...},  \label{R}
\end{equation}
where the sum over $k$ and $l$ includes all resonances, and $b\left(
m\rightarrow n\right) $ is the branching ratio for the decay process $%
m\rightarrow n$. If the decay process $m\rightarrow n$ does not take place,
the branching ratio $b\left( m\rightarrow n\right) $ is taken to be zero.
The inclusion of the resonances is a very important effect, since the
contributions from the decays are quite substantial, especially at large
temperatures and densities. Eqs. (\ref{R}) may be also used in a slightly
different form, namely 
\begin{equation}
R=\frac{n_{i}+\sum_{k}b\left( k\rightarrow i\right) n_{k}+\sum_{kl}b\left(
k\rightarrow l\right) b\left( l\rightarrow i\right) n_{k}+...}{%
n_{j}+\sum_{k}b\left( k\rightarrow j\right) n_{k}+\sum_{kl}b\left(
k\rightarrow l\right) b\left( l\rightarrow j\right) n_{k}+...},  \label{R1}
\end{equation}
where 
\begin{equation}
n_{i}=\frac{g_{i}}{2\pi ^{2}}\int_{0}^{\infty }\frac{p^{2}\ dp}{\exp \left[
\left( E_{i}-\mu _{chem}^{B}B_{i}-\mu _{chem}^{S}S_{i}-\mu
_{chem}^{I}I_{i}\right) /T_{chem}\right]+ \epsilon} \, .  \label{ni}
\end{equation}
To stress the fact that we calculate the ratios at the chemical freeze-out,
we have marked the temperature and the chemical potentials with an
additional label (in this way we also make a connection with the notation
used in Refs. \cite{MichFlorBron-Scal} and \cite{FlorBron}).

\begin{figure}[t]
\epsfysize=8cm
\par
\begin{center}
\mbox{\epsfbox{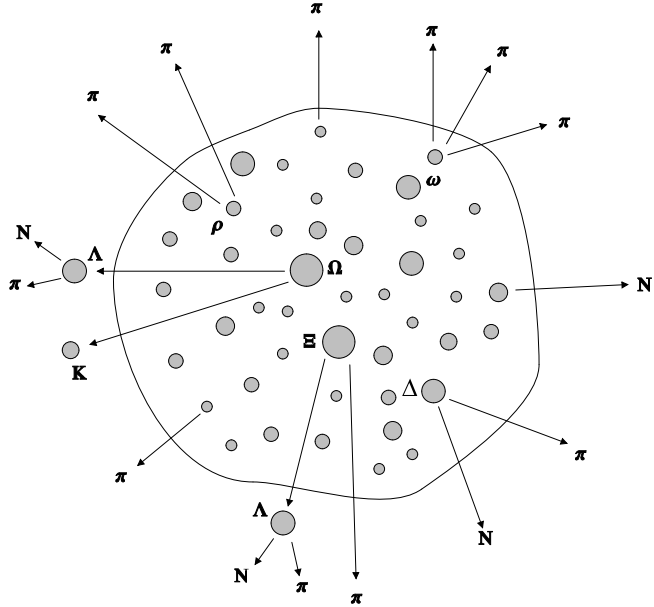}}
\end{center}
\caption{{\small At the chemical freeze-out, the hadron gas consists of stable
hadrons (with respect to strong interacions) and all hadronic
resonances.
The observed multiplicities obtain contributions from the
primary particles (present in the fireball), and secondary particles
coming from the strong decays of the resonances. In addition, the
electro-weak decays should be reconstructed. In this way the yields of
strange particles can be obtained. }}
\label{ratios}
\end{figure}

\subsubsection{i) \underline{ Isospin Symmetry}}

In our approach we take into consideration all charged (isospin) states of
hadrons separately. The non-zero value of the fitted chemical potential $\mu
^{I}$ leads to the splitting between the multiplicities of the hadrons
belonging to a given isospin multiplet. Consequently, the ratios of such
hadrons may be included in our analysis. For example, we take into
consideration the $\pi ^{+}/\pi ^{-}$ ratio measured at SPS and RHIC. The
typical values of $\mu ^{I}$ are of the order of 10 MeV, thus the inclusion
of the isospin chemical potential represents an important effect (only at
RHIC $\mu ^{I}$ is practically zero).

A correct treatment of the different charged states requires that the
branching ratios supplied by the Review of the Particle Physics
\cite{PDG} must be supplemented by the calculation of the probability
of the transitions between such states. Since the strong interactions
are invariant under rotations in the isospin space, the matrix
elements needed for the two-body strong decays may be obtained
directly from the Clebsch-Gordan coefficients. \footnote{For the
three-body decays an additional averaging is needed, unless the full
matrix element describing the decay is known.} For example, we can
consider the strong $\Delta \longrightarrow \pi N$ decay. Practically,
this is the only decay channel of $\Delta$, so its branching ratio is
100\%. In order to include different charged states of $\Delta ,\,\pi
$ and the nucleon, $N=(p,n)$, we combine the isospin states 1 and 1/2
into the isospin states 3/2:
\begin{eqnarray*}
\Delta ^{++} &\longrightarrow &\,\left\langle 1\,1,\frac{1}{2}\,\frac{1}{2}%
\,|\frac{3}{2}\,\frac{3}{2}\right\rangle \,(\pi ^{+}+p)=\pi ^{+}+p, \\
\Delta ^{+} &\longrightarrow &\,\left\langle 1\,1,\frac{1}{2}\,-\frac{1}{2}%
\,|\frac{3}{2}\,\frac{1}{2}\right\rangle \,\left( \pi ^{+}+n\right)
+\left\langle 1\,0,\frac{1}{2}\,\frac{1}{2}\,|\frac{3}{2}\,\frac{1}{2}%
\right\rangle \,\left( \pi ^{0}+p\right) \\
&=&\frac{1}{3}\left( \pi ^{+}+n\right) +\frac{2}{3}\left( \pi ^{0}+p\right) ,
\\
\Delta ^{0} &\longrightarrow &\,\left\langle 1\,-1,\frac{1}{2}\,\frac{1}{2}%
\,|\frac{3}{2}\,-\frac{1}{2}\right\rangle \,\left( \pi ^{-}+p\right)
+\left\langle 1\,0,\frac{1}{2}\,-\frac{1}{2}\,|\frac{3}{2}\,-\frac{1}{2}%
\right\rangle \,\left( \pi ^{0}+n\right) \\
&=&\frac{1}{3}\left( \pi ^{-}+p\right) +\frac{2}{3}\left( \pi ^{0}+n\right) ,
\\
\Delta ^{-} &\longrightarrow &\,\left\langle 1\,-1,\frac{1}{2}\,-\frac{1}{2}%
\,|\frac{3}{2}\,-\frac{3}{2}\right\rangle \,(\pi ^{-}+n)=\pi ^{-}+n.
\end{eqnarray*}

\subsubsection{ii) \underline{ Experimental Branching Ratios}}

The case $\Delta \longrightarrow \pi N$ is very comfortable, since
only a single decay channel is present and the branching ratio is
obviously well known. In majority of the cases we deal with, different
decay channels appear and their properties (branching ratios) are
sometimes not well known.  As a rule we disregard all decays with the
branching ratios smaller than 1\%. In addition, if the decay channels
are described as {\it dominant}, {\it large, seen}, or {\it possibly
seen, }we always take into account the most important channel. If two
or more channels are described as equally important, we take all of
them with the same weight. For example $f_{0}(980)$ decays into $\pi
\pi $ (according to \cite{PDG} this is the {\it dominant} channel) and
$K \overline{K}$ (according to \cite{PDG} this is the {\it seen}
channel). In our approach, according to the rules stated above we
include only the process $f_{0}(980) \longrightarrow \pi \pi
$. Similarly, $a_{0}(1450)$ has three decay channels: $\eta \pi $
({\it seen}), $\pi \eta ^{\prime }(958)$ ({\it seen}), and $K
\overline{K}$ (again {\it seen}). In this case we include all three
decay channels with the weight (branching ratio)\ 1/3.

\begin{table}[t]
\begin{center}
\begin{tabular}{|l|l|l|l|}
\hline
$\Delta (1600)$ decay modes & branching ratio & $
\begin{tabular}{l}
averaged \\ 
branching ratio
\end{tabular}
$ & 
\begin{tabular}{l}
rescaled \\ 
branching ratio
\end{tabular}
\\ \hline \hline
$N\pi $ & \multicolumn{1}{|r|}{10-25 \%} & 17.5 \% & 17.5 \% \\ \hline
$N\pi \pi $ & \multicolumn{1}{|r|}{75-90 \%} & 82.5 \% & 82.5 \% \\ 
\quad $\Delta \pi $ & \multicolumn{1}{|r|}{40-70 \%} & \multicolumn{1}{|c|}{
55.0 \%} & \multicolumn{1}{|c|}{50.4 \%} \\ 
\quad $N\rho $ & \multicolumn{1}{|r|}{$<$ 25 \%} & \multicolumn{1}{|c|}{12.5
\%} & \multicolumn{1}{|c|}{11.5 \%} \\
\quad $N(1440)\pi $ & \multicolumn{1}{|r|}{10-35 \%} & \multicolumn{1}{|c|}{
22.5 \%} & \multicolumn{1}{|c|}{20.6 \%} \\ \hline
$N\gamma $ & \multicolumn{1}{|r|}{0.001-0.02 \%} & \multicolumn{1}{|c|}{---}
& \multicolumn{1}{|c|}{---} \\ 
\quad $N\gamma ,\,$helicity=1/2 & \multicolumn{1}{|r|}{0.0-0.02 \%} & 
\multicolumn{1}{|c|}{---} & \multicolumn{1}{|c|}{---} \\ 
\quad $N\gamma ,\,$helicity=3/2 & \multicolumn{1}{|r|}{0.001-0.005 \%} & 
\multicolumn{1}{|c|}{---} & \multicolumn{1}{|c|}{---} \\ \hline
\end{tabular}
\end{center}
\caption{{\small The branching ratios for the decays of
$\Delta$(1600). The experimental information (second column) is
averaged (third column) and rescaled (fourth column) in order to
achieve the correct normalization. Similar procedure is also applied
for other decays whose branching ratios are poorly known. }}
\label{D1600}
\end{table}

Another difficulty is that the branching ratios are not given exactly
(instead of one value, the whole range of acceptable values is given)
and the sum of the branching ratios may differ significantly from
1. In this case we take the mean values of the branching ratios. Since
we require that their sum is properly normalized, sometimes we are
forced to rescale all the mean values in such a way that their sum is
indeed 1. To illustrate this problem we present our analysis of the
decays of the $\Delta (1600)$ resonance in Table \ref{D1600}. Below,
as an example we give the final branching ratios of the
$\Delta(1600)^{++}$ resonance:
\begin{eqnarray*}
\Delta (1600)^{++} &\longrightarrow &0.175\,\left(\pi ^{+}+p\right) \\
&&+\, 0.504\left[\, \frac{2}{5}\left( \Delta ^{+}+\pi ^{+}\right) +\frac{3}{5}%
\left(\Delta ^{++}+\pi ^{0}\right)\right]  \\
&&+\, 0.115\,\left(\, p+\rho ^{+}\right) \\
&&+\, 0.206\,\left[N(1400)^{+}+\pi ^{+}\right].
\end{eqnarray*}
We note that the correct normalization of the branching
ratios is crucial for the fulfillment of the conservation laws, and
consequently for the determination of the optimal thermodynamic
parameters.

\begin{figure}[h]
\epsfysize=6cm
\par
\begin{center}
\mbox{\epsfbox{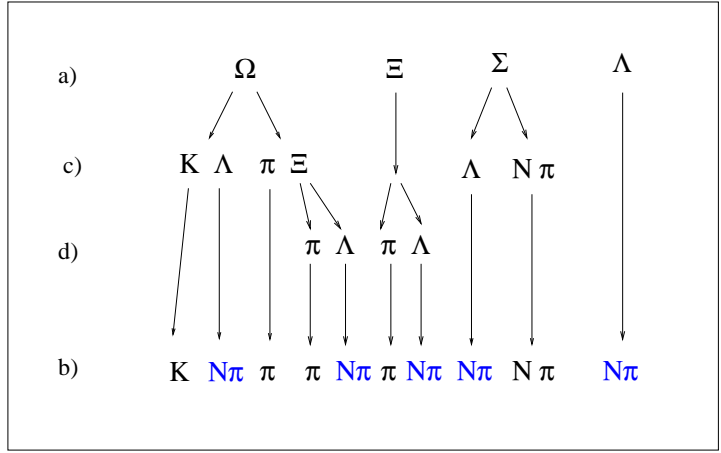}}
\end{center}
\caption{{\small A schematic view of the weak decays of strange baryons.}}
\label{weakd}
\end{figure}

\subsubsection{iii) \underline{ Weak Decays}}

An important part of the thermal analysis is also the correct
treatment of the contributions from the electro-weak decays. For
example, the final (measured) multiplicity of pions is modified by the
decays of $\Lambda ,\Sigma $ and $\Xi .$ Since the abundances of
$\Lambda ,\Sigma $ and $\Xi $ in a hot and dense matter are not
negligible, their decays modify the final pion multiplicity. To study
such effects, we do our calculations usually in three different
ways. In the first version we include{\it \ all contributions } from
the weak decays. In the second version we assume that there is {\it no
feedback} from the weak decays -- this case corresponds to the ideal
situation when all weak-decay channels can be experimentally
disentangled.  Finally, in the third version we assume that the
feedback from the weak decays is at the level of 50\%. Here we follow
the concept of Ref. \cite {PBMHS}.

To be more precise, in our description of the weak decays the labels
$a,b,c$ and $d$ are introduced (see Fig. \ref{weakd}).  The label $a$
specifies the particle multiplicity which includes the two
contributions: the first one from the primordial particles, and the
second one from all {\it strong-decay} channels. At the $ a $-level
our system includes: $\pi ,\eta ,K,N,\Lambda ,\Sigma ,\Xi $ and $
\Omega $. The label $c$ characterizes the abundances corrected for the
$ \Omega $ and $\Sigma $ decays, whereas label $d$ means that the
decays of $ \Xi $ are also included. At the $d$-level we deal with:
$\pi ,\eta ,K,N$ and $\Lambda $. The final abundances of pions, kaons
and nucleons (including all electro-weak decays) are denoted by the
label $b$.

\subsubsection{iv) \underline{Resonance Mass Spectrum}}

In Refs. \cite{hag2,brohag1,brohag2} the arguments have been presented
that the mass spectra of baryons and mesons behave differently: the
baryon spectrum grows more rapidly than the meson spectrum. As a consequence,
the {\it Hagedorn temperature} (i.e., a scale describing the exponential
growth of the spectrum) is different for baryons and mesons. With the use
of a simple exponential formula, the spectra of baryons and mesons
may be well parameterized as follows:
\begin{equation}
\rho_{B,M}(m) = A_{B,M} \exp \left[
{m \over T^{B, M}_H } \right],
\label{rBM}
\end{equation}
with the {\it baryon Hagedorn} temperature
\begin{equation}
T^B_H = 186 \, \hbox{MeV},
\label{tbh}
\end{equation}
the {\it meson Hagedorn} temperature
\begin{equation}
T^M_H = 311 \, \hbox{MeV},
\label{tmh}
\end{equation}
and the normalization constants: $A_B = $ 4.41 GeV and $A_M =$
0.11 GeV \cite{wb}.

\subsection{Sketch of $\protect\chi ^{2}$ Method}

The optimal values of the temperature $T_{chem}$ and the baryon chemical
potential $\mu _{chem}^{B}$ are fitted by minimizing the expression

\begin{equation}
\chi ^{2}=\sum_{k=1}^{n}\frac{\left( R_{k}^{\exp }-R_{k}^{therm}\right) ^{2}%
}{\sigma _{k}^{2}}, 
\label{chi2}
\end{equation}
where $R_{k}^{\exp }$ is the $k$th measured ratio, $\sigma _{k}$ is the
corresponding error, and $R_{k}^{therm}$ is the same ratio determined from
the thermal model. The total number of different ratios included in the
analysis is denoted by $n$.

Introducing the short-hand notation $\alpha _{1}=T_{chem}$ and $\alpha
_{2}=\mu _{chem}^{B}$ we may write 
\begin{equation}
\chi ^{2}\left( \alpha \right) =\chi ^{2}\left( \alpha _{\min }\right)
+\left( \alpha -\alpha _{\min }\right) ^{T}V^{-1}\left( \alpha -\alpha
_{\min }\right) ,
\label{chi2alfa} 
\end{equation}
where $\alpha =\left( \alpha _{1},\alpha _{2}\right) $, $\alpha _{\min }$ is
the optimal pair of the parameters, and $V$ is the variance matrix of the
parameters $\alpha $. We note that $\chi ^{2}\left( \alpha \right) $ has a
minimum at $\alpha =\alpha _{\min }$, so the first derivatives of $\chi
^{2}\left( \alpha \right) $ vanish at this point. If $F\left( \alpha \right) 
$ is some function of the fitted parameters $\alpha $, the variance of $F$
is given by 
\begin{equation}
\left( \Delta F\right) ^{2}=\sum_{mn}\frac{\partial F}{\partial \alpha _{m}}%
\frac{\partial F}{\partial \alpha _{n}}V_{mn}.  \label{df2}
\end{equation}
In the special cases $F=\alpha _{1}$ and $F=\alpha _{2}$ , Eq. (\ref{df2})
can be used to make an estimate of the errors of the fitted temperature and
baryon chemical potential 
\begin{equation}
\left( \Delta T_{chem}\right) ^{2}=V_{11},\qquad \left( \Delta \mu
_{chem}^{B}\right) ^{2}=V_{22}.  \label{v11v22}
\end{equation}
Equation (\ref{v11v22})\ will be used in our analysis with the coefficients $%
V_{11}$ and $V_{22}$ determined numerically.

\subsection{Thermodynamic Characteristics of Freeze-Out}

It is very interesting to know the thermodynamic properties of the
hadronic matter at freeze-out. Of the particular interest is the
energy density of the hadronic system. Its closeness to the critical
energy density for the deconfinement phase transition (found by the
Monte-Carlo simulation of QCD \ on the lattice \cite{Karsch}) may
indicate that such a phase transition indeed took place in the
ultra-relativistic heavy-ion collisions. In our approach the knowledge
of the temperature and the chemical potentials allows us to calculate
all intensive thermodynamic quantities. In particular, we calculate
the energy density
\begin{equation}
\varepsilon = \sum_i \varepsilon_i
=\sum_{i}g_{i}\int d^{3}p\ E_{i}\left( p\right) \ f_{i}\left(
p\right) ,  
\label{epsi}
\end{equation}
pressure 
\begin{equation}
P=\sum_i P_i =
\sum_{i}g_{i}\int d^{3}p\ \frac{p^{2}}{3\ E_{i}\left( p\right) }%
f_{i}\left( p\right) ,  \label{pres}
\end{equation}
and the baryon, strangeness, and isospin densities 
\begin{equation}
\rho _{B} = \sum_{i} g_{i} \int d^{3}p \, B_{i} \, f_{i}(p), 
\label{rhoBe}
\end{equation}
\begin{equation}
\rho _{S} = \sum_{i} g_{i} \int d^{3}p \, S_{i} \, f_{i}(p), 
\end{equation}
\begin{equation}
\rho _{I} = \sum_{i} g_{i} \int d^{3}p \, I_{i} \, f_{i}(p). 
\end{equation}
The entropy density is determined from the Gibbs identity
\begin{equation}
s=\frac{\varepsilon +P-\mu _{B}\rho _{B}-\mu _{S}\rho _{S}-\mu _{I}\rho _{I}%
}{T}.  
\label{entropy}
\end{equation}

It is interesting to discuss the case of the classical statistics
separately (this is achieved by taking the limit $\epsilon \rightarrow
0$ in the equilibrium distribution functions (\ref{fip})). In this
case the thermodynamic quantities can be expressed in terms of the
modified Bessel functions
\begin{equation}
K_{n}(x)=\frac{2^{n}n!}{(2n)!}x^{-n}\int_{x}^{\infty}d \tau \left (
\tau^{2}-x^{2} \right )^{n-1/2}  e^{-\tau}. 
\label{Kn}
\end{equation}
In particular, the particle densities can be found from expression
\begin{equation}
n_{i}=\frac{1}{2\pi^{2}}e^{\mu_i /T}m_{i}^{2}TK_{2}\left( \frac{m_{i}}{T}%
\right) ,  \label{claflow}
\end{equation}
and the corresponding energy density and the pressure are determined by
the following two equations
\begin{eqnarray}
\varepsilon _{i}+P_{i}&=&\frac{1}{2\pi^{2}}e^{\mu_i /T}m_{i}^{3}TK_{3}\left( 
\frac{m_{i}}{T}\right)  \\
P_{i}&=&\frac{1}{2\pi^{2}}e^{\mu_i/T}m_{i}^{2}T^{2}
K_{2}\left( \frac{m_{i}}{T}\right) .  \label{claepsp}
\end{eqnarray}
One can notice that Eqs. (\ref{claflow}) and (\ref{claepsp}) yield the
equation of state of the relativistic ideal gas of classical particles
($k_{B}=1$)
\begin{equation}
P=nT.  
\label{claEOS}
\end{equation}
We note that it is the same as the equation of state of the
non-relativistic ideal gas.  

It has been pointed out by Cleymans and Redlich \cite{CleyRed}, that
for a variety of different colliding systems the average energy per
hadron at the chemical freeze-out is very close to 1 GeV 
\begin{equation}
r = {\varepsilon \over n} = {\sum_i \varepsilon_i \over
\sum_i n_i} \approx 1 \, \hbox{GeV}.
\end{equation}
In our calculations we check this relation. In addition, we calculate
separately the average energy of baryons and mesons defined as
\begin{equation}
r_B = {\varepsilon_B \over n_B} = {\sum\limits_{baryons} \varepsilon_i \over
\sum\limits_{baryons} n_i}\,\,,
\hspace{1cm}
r_M = {\varepsilon_M \over n_M} = {\sum\limits_{mesons} \varepsilon_i \over
\sum\limits_{mesons} n_i}.
\end{equation}

\section{Modifications of Thermal Analysis}

\subsection{Finite-Size Corrections}
\label{fsc}

The particle densities defined by the ideal-gas expression correspond to the
thermodynamic limit, i.e., they are calculated in the limit $%
V\longrightarrow \infty ,\,N\longrightarrow \infty $, $N/V=\,$const. In
realistic situations, the hadronic systems have finite volumes and the
approach based on Eq. (\ref{ni}) may be not sufficiently accurate. To
include the effects of the restricted volume, one can use the modified
versions of Eq. (\ref{ni}). In practice, such modifications are known only
for simple geometries. In the special case of the spherical symmetry, when
the hadronic system forms a fireball of radius $R_{s}$, \ a correction of
Eq. (\ref{ni}) may be achieved by the change of the level density \cite{JMZ}
\begin{equation}
d^{3}p\longrightarrow d^{3}p\left( 1-\frac{3\pi }{4pR_{s}}+\frac{1}{%
(pR_{s})^{2}}+...\right) .  \label{modd3p}
\end{equation}

\noindent Hence, the particle densities may be calculated from the formula

\begin{equation}
n_{i}=\frac{g_{i}}{2\pi ^{2}}\int_{0}^{\infty }dp\left( p^{2}-\frac{3\pi p}{%
4R_{s}}+\frac{1}{R_{s}^{2}}+...\right) \ \left[ \exp \left( \frac{E_{i}-\mu
_{i,chem}}{T_{chem}}\right) + \epsilon \right] ^{-1}  \label{modni}
\end{equation}
Clearly, for very large radii, $R_{s}\longrightarrow \infty ,$
Eq. (\ref {modni}) is reduced to the standard formula (\ref{ni}).  The
practical calculations \cite{PBM-AGS,PBM-SPS} indicate that the
finite-size corrections affect mainly the absolute densities. The
particle ratios are not changed much due to this effect.

\subsection{Excluded-Volume (van der Waals) Corrections}
\label{exvol}

The excluded-volume corrections account for the finite volumes of
hadrons.  In a very dense hadronic matter the particles with finite
size may overlap.  This leads to strong repulsive forces and the
simple approach based on Eqs. ( \ref{Ni}) and (\ref{fip}) should be
modified (by definition Eqs. (\ref{Ni}) and (\ref{fip}) describe
non-interacting particles). Of course, the overlapping of hadrons may
be the first step towards the deconfinement phase transition and
creation of the quark-gluon plasma. In any case, however, the
repulsive part of the nuclear force at short distances should be
included in the realistic description of the hadron thermodynamics.

A fully consistent (from the thermodynamic point of view) method to
include the excluded volume corrections was introduced by Yen,
Gorenstein, Greiner and Yang \cite{YGGY}. In their approach one
calculates the modified pressure $ \widetilde{P}$ defined by

\begin{equation}
\widetilde{P}(T,\mu _{1},\mu _{2},...)=P(T,\widetilde{\mu }_{1},\widetilde{%
\mu }_{2},...),  \label{presmod}
\end{equation}
where $P$ is the pressure of the ideal gas, as defined by
Eq. (\ref{pres}), and $\widetilde{\mu }_{i}$ are the modified chemical
potentials

\begin{equation}
\widetilde{\mu }_{i}=\mu _{i}-v_{i}\,\widetilde{P}(T,\mu _{1},\mu
_{2},...).\quad  \label{muimod}
\end{equation}
In Eq. (\ref{muimod}) $v_{i}$ is the particle eigenvolume
\cite{YGGY}. We note that Eq. (\ref {presmod})\ is a non-linear
equation for $\widetilde{P}$, which can be solved by the iterative
method. The particle densities $\widetilde{n}_{i}$ are the derivatives
of $\widetilde{P}$ with respect to the chemical potentials $\mu _{i}$
\begin{equation}
\widetilde{n}_{i}\equiv \frac{\partial \widetilde{P}}{\partial \mu _{i}}%
=\sum_{j}\frac{\partial P}{\partial \widetilde{\mu }_{j}}\frac{\partial
\widetilde{\mu }_{j}}{\partial \mu _{i}}=\sum_{j}n_{j}\left( \delta
_{ji}-v_{j}\,\widetilde{n}_{i}\right) ,  \label{nit}
\end{equation}
which gives

\begin{equation}
\widetilde{n}_{i}(T,\mu _{i})\equiv \frac{n_{i}(T,\widetilde{\mu }_{i})}{%
1+\sum_{j}v_{j}\,n_{j}(T,\widetilde{\mu }_{j})}.  \label{nit1}
\end{equation}
We observe that the modified densities are {\it smaller} than the initial
densities. The reasons for this are twofold: firstly, \ the denominator
appearing on the right-hand side of Eq. (\ref{nit1}) is always larger than
one; secondly, the densities $n_{i}$ are calculated at smaller values of the
chemical potential, since we have always $\widetilde{\mu }_{i}\leq \mu _{i}$.

For our investigation of the relative particle yields, it is important to
realize that the denominator in Eq. (\ref{nit1}) is the same for all hadron
species. Hence, it cancels in the particle ratios. In addition, for the
classical (Boltzmann) statistics and identical particle volumes of all
hadrons ($v_{i}=v$), the modifications of the chemical potential factorize,
and again cancel in the ratios. {\it Thus, in most cases the excluded volume
corrections do not affect the thermal analysis of the particle ratios and
have no impact on the fitted value of the optimal temperature and baryon
chemical potential}

\begin{equation}
\frac{\widetilde{n}_{i}(T,\mu _{i})}{\widetilde{n}_{j}(T,\mu _{i})}\approx
\frac{n_{i}(T,\mu _{i})}{n_{i}(T,\mu _{i})}.  \label{ntintj}
\end{equation}
Note, however, that the actual particle densities should be calculated from
Eq. (\ref{muimod}) with the modified chemical potential $\widetilde{\mu }%
_{i} $.

The case of the Boltzmann statistics is also interesting, since it leads
directly to the van der Waals equation of state. To see this feature, we
first write 
\begin{equation}
\widetilde{P}=\exp \left[ -\frac{v\widetilde{P}}{T}\right] P.  \label{wdw1}
\end{equation}
A derivative of Eq. (\ref{wdw1}) with respect to the chemical potentials
yields 
\begin{equation}
\widetilde{P}\left( 1-v\widetilde{n}\right) =\widetilde{n}T,\qquad 
\widetilde{n}=\sum_{i}\widetilde{n}_{i},  \label{wdw2}
\end{equation}
which is a typical excluded volume correction.

\subsection{Corrections for Undersaturation of Strangeness}
\label{cfuos}

The chemical equilibrium between strange and non-strange particles is
more difficult to achieve because the production of $\bar{s}s$ pairs
proceeds usually at a slower rate than the production of $\bar{u}u$
and $\overline{d} d $ pairs. Moreover, there are no strange quarks at
the beginning of the collision process. In Ref. \cite{RafLetTou} \
Rafelski has introduced an extra parameter to account for possible
undersaturation of strangeness. In that approach \ one
calculates the hadronic densities from the formula
\begin{equation}
n_{i}=\frac{g_{i}}{2\pi ^{2}}\int_{0}^{\infty }dpp^{2}\ \left[ \gamma
^{-s_{i}}\exp \left( \frac{E_{i}-\mu _{i,chem}}{T_{chem}}\right) + \epsilon \right]
^{-1},  \label{nis}
\end{equation}
where $\gamma $ is the strangeness saturation factor ($0<\gamma \leq 1$) and 
$s_{i}$ is the number of valence strange quarks in the $i$th hadron.

\section{Thermal Model in this Work}

In the following Sections we apply the thermal model to describe the
particle ratios measured in different experiments with
ultra-relativistic heavy-ions. We use the simplest version of the
model, and usually neglect the corrections discussed in Sections \ref{fsc} --
\ref{cfuos}. In other words, we assume the full chemical and thermal
equilibrium, and use the ideal-gas formulas. The role of the
corrections defined in Sections \ref{fsc} -- \ref{cfuos} is discussed
in special cases. Our method of dealing with the resonance decays was
presented in detail in Section \ref{dor}.  The results of our
calculations are denoted in the text by the label TM.

%%%%%%%%%%%%%%%%%%%%%%%%%%%%%%%%%%%%%%%%%%%%%%%%%%%%%%%%%%%%%%%%%%

\chapter{Reactions with Si and S Beams}

In this Chapter we use our implementation of the thermal model to
analyze some of the first experiments with the ultra-relativistic
heavy-ions: the experiments with the ${}^{28}$Si beams at BNL AGS, and
the experiments with $ ^{32}$S beams at CERN SPS. The Alternating
Gradient Synchrotron at BNL accelerated ${}^{28}$Si ions to the energy
of 14.6 GeV per nucleon (in 1986), whereas the Super Proton
Synchrotron at CERN accelerated $^{32}$S to the energy of 200 GeV per
nucleon (in 1987 and later in 1990). For many years the results from
sulphur-induced reactions were the main source of the data on hadron
production at true ultra-relativistic energies. 

\section{Si + Au(Pb) Collisions at BNL\ AGS}

The heavy-ion collisions with Si beams were analyzed in a thermal
model by Braun-Munzinger, Stachel, Wessels, and Xu \cite{PBM-AGS} (for
Au and Pb targets), and later by Cleymans, Elliot, Satz, and Thews
\cite{CleyEllSa} (for Au targets). In this Section we are going to
compare their results with our analysis of the particle ratios.

In Ref. \cite{PBM-AGS} the finite-size and the excluded-volume
corrections were included, and the particle ratios were calculated for
two different temperatures, $T_{chem}$ = 120 MeV and $T_{chem}$ = 140
MeV, and the fixed baryon chemical potential $\mu^B_{chem}$ = 540 MeV.
The range of temperatures considered in \cite{PBM-AGS} was motivated
by the experimental spectrum of the $\Delta$(1232) resonance. The
value of the baryon chemical potential was constrained by the measured
pion to nucleon ratio. For a given temperature and baryon chemical
potential the strangeness chemical potential was adjusted to give zero
net strangeness (compare Eq. (\ref{s0})). On the other hand, the
isospin chemical potential was neglected.  The results of
Ref. \cite{PBM-AGS} are shown in Table \ref{SiAuRatio} in the fifth
(BM1) and seventh (BM2) column: one can see that the overall agreement
with the data (second column) has been achieved, but the statistical
significance of this result is rather poor, which is indicated by the
very large values of $\chi^2$. In the two discussed cases one finds:
$\chi^2/n=8.5$ and $\chi^2/n=12.2$.

%\newpage

We have reanalyzed the experimental data (as compiled in
\cite{PBM-AGS}) in our model assuming the full chemical equilibrium
and neglecting the finite-size and excluded-volume corrections. The
results of our fit are shown in the third column (TM). We have found a
slightly better fit with $\chi^2/n=6.3$ for $T_{chem}$ = 136 MeV and
$\mu^B_{chem}$ = 593 MeV. Still, the value of $\chi^2/n$ is not
satisfactory. As expected, using the values of Ref. \cite{PBM-AGS} for
$T_{chem}$ and $\mu^B_{chem}$ as an input in our model (without the
fitting procedure), we find the ratios corresponding to higher values
of $\chi^2/n$. These results are shown in the fourth and sixth column
denoted by TM1 and TM2, respectively. We note that our determination
of the strange chemical potential agrees with the result of
\cite{PBM-AGS} in these two cases.

\begin{table}
\begin{center}
\begin{tabular}{|l|l|l||l|l||l|l|}
\hline
{\bf Si+Au(Pb)} & experiment & TM & TM 1 & BM 1 & TM 2 & BM 2 \\ \hline\hline
$T_{chem}$ [MeV] &  & \multicolumn{1}{|r||}{{\bf 136}$\pm {\bf 2}$} &
\multicolumn{1}{|r|}{120} & \multicolumn{1}{|r||}{{\bf 120}} &
\multicolumn{1}{|r|}{ 140} & \multicolumn{1}{|r|}{{\bf 140}} \\ \hline
$\mu _{chem}^{B}$ [MeV] &  & \multicolumn{1}{|r||}{{\bf 593}$\pm {\bf 9}$} &
\multicolumn{1}{|r|}{540} & \multicolumn{1}{|r||}{{\bf 540}} &
\multicolumn{1}{|r|}{ 540} & \multicolumn{1}{|r|}{{\bf 540}} \\ \hline
$\mu _{chem}^{S}$ [MeV] &  & \multicolumn{1}{|r||}{152} &
\multicolumn{1}{|r|}{109} & \multicolumn{1}{|r||}{108} & \multicolumn{1}{|r|}{
135} & \multicolumn{1}{|r|}{135} \\ \hline
$\mu _{chem}^{I}$ [MeV] &  & \multicolumn{1}{|r||}{-14} &
\multicolumn{1}{|r|}{-11} & \multicolumn{1}{|r||}{0} & \multicolumn{1}{|r|}{
-13} & \multicolumn{1}{|r|}{0} \\ \hline
$\chi ^{2}/n$ &  & \multicolumn{1}{|r||}{{\bf 6.3}} & \multicolumn{1}{|r|}{9.3}
& \multicolumn{1}{|r||}{{\bf 8.5}} & \multicolumn{1}{|r|}{22.1} &
\multicolumn{1}{|r|}{{\bf 12.2}} \\ \hline\hline
$\pi /(p+n)$ & 1.05(5) & 1.20 & 1.36 & 1.29 & 1.45 & 1.34 \\ \hline
$\bar{p}/p$ & 4.5(5)$\cdot 10^{-4}$ & 3.6$\cdot 10^{-4}$ & 2.1$\cdot 10^{-4}$
& 1.47$\cdot 10^{-4}$ & 9.0$\cdot 10^{-4}$ & 5.8$\cdot 10^{-4}$ \\ \hline
$K^{+}/\pi ^{+}$ & 0.19(2) & 0.23 & 0.23 & 0.23 & 0.24 & 0.27 \\ \hline
$K^{-}/\pi ^{-}$ & 3.5(5)$\cdot 10^{-2}$ & 3.3$\cdot 10^{-2}$ & 3.9$\cdot
10^{-2}$ & 5.0$\cdot 10^{-2}$ & 4.3$\cdot 10^{-2}$ & 6.2$\cdot 10^{-2}$ \\
\hline
$K_{s}^{0}/\pi ^{+}$ & 9.7(15)$\cdot 10^{-2}$ & 14$\cdot 10^{-2}$ & 14$\cdot
10^{-2}$ & 14$\cdot 10^{-2}$ & 15$\cdot 10^{-2}$ & 16$\cdot 10^{-2}$ \\
\hline
$K^{+}/K^{-}$ & 4.4(4) & 5.5 & 4.6 & 4.6 & 4.4 & 4.3 \\ \hline
$\Lambda /(p+n)$ & 8.0(16)$\cdot 10^{-2}$ & 7.2$\cdot 10^{-2}$ & 7.6$\cdot
10^{-2}$ & 9.5$\cdot 10^{-2}$ & 8.6$\cdot 10^{-2}$ & 11$\cdot 10^{-2}$ \\
\hline
$\overline{\Lambda}/\Lambda $ & 2.0(8)$\cdot 10^{-3}$ & 2.3$\cdot 10^{-3}$ & 1.0$%
\cdot 10^{-3}$ & 0.88$\cdot 10^{-3}$ & 4.4$\cdot 10^{-3}$ & 3.7$\cdot
10^{-3} $ \\ \hline
$\phi /\left( K^{+}+K^{-}\right) $ & 1.34(36)$\cdot 10^{-2}$ & 2.89$\cdot
10^{-2}$ & 2.4$\cdot 10^{-2}$ & 2.4$\cdot 10^{-2}$ & 3.5$\cdot 10^{-2}$ & 3.6%
$\cdot 10^{-2}$ \\ \hline
$\Xi ^{-}/\Lambda $ & 0.12(2) & 0.05 & 0.056 & 0.064 & 0.059 & 0.072 \\
\hline
\end{tabular}
\end{center}
\caption{{\small Thermal analysis of the particle ratios measured in
Si + Au(Pb) reactions. The experimental ratios are used in the form
prepared by Braun-Munzinger et al. in Ref. \cite{PBM-AGS}. The optimal
values of the thermodynamic parameters found in our approach are
presented in the third column (TM).  The results of
Ref. \cite{PBM-AGS} are presented in the fifth and seventh column
(denoted by BM1 and BM2, respectively). For comparison the results of
our model for the same input values of $\,T_{chem}$ and $\mu^B_{chem}$
are shown in the fourth and sixth column (TM1 and TM2,
respectively). In our calculations (TM,TM1,TM2) we include the weak
decays at the level of 50\%. }}
\label{SiAuRatio}
\end{table}

In Table \ref{SiAu} we show the values of the thermodynamic parameters
which follow from our analysis of the particle ratios.  Our best fit
gives the energy density $\varepsilon$ = 0.6 GeV/fm$^3$, and the
baryon number density $\rho_B$ = 0.34 fm$^{-3}$ (we use here
Eqs. (\ref{epsi}) and (\ref{rhoBe})). In all considered cases the
pressure satisfies the condition $P = (n_M+n_B)\, T$, which indicates
that the matter at the chemical freeze-out behaves like an ideal {\it
classical} gas (compare Eq. (\ref{claEOS})). In this situation the
excluded-volume corrections cancel in the particle ratios and do not
affect the optimal values of the temperature and the baryon chemical
potential (as discussed in the end of Section \ref{exvol}).
The classical form of the equation of state is a consequence of the
fact that most of the particles included in the thermal approach are
heavy resonances, which are well described by the classical
distribution functions.  The important role of heavy resonances in the
thermal analysis is reflected also by the fact that the final density
of pions (resulting from the decays of resonances) is much higher than the
primordial density. Initially, the pion density is 0.09 fm$^{-3}$.
With the inclusion of the strong decays of the resonances, the pion
density increases up to 0.35 fm$^{-3}$.  We also note that the
contribution from the decays of the $\rho$-mesons to the final pion
density is at the level of 15\%.  This means that most of the extra
pions is produced by the decays of heavier resonances.  The last raw
in Table \ref{SiAu} shows the Cleymans-Redlich ratio calculated for
three different choices of $T_{chem}$ and $\mu^B_{chem}$. In all cases 
we observe that $r \approx$ 1 GeV. It is interesting to notice,
however, that $r_B$ is significantly larger than $r_M$. This feature
reflects a different behavior of the mass spectra of baryons and
mesons, as displayed by Eqs. (\ref{rBM}) - (\ref{tmh}).

\begin{table}
\begin{center}
\begin{tabular}{|l|l|l|l|}
\hline
{\bf Si+Au(Pb)} & TM & TM 1 & TM 2 \\ \hline \hline
$T_{chem}$ [MeV] & \multicolumn{1}{|r|}{136} & 120 & 140 \\ \hline
$\mu _{chem}^{B}$ [MeV] & \multicolumn{1}{|r|}{593} & 540 & 540 \\
\hline
$\mu _{chem}^{S}$ [MeV] & \multicolumn{1}{|r|}{152} & 109 & 135 \\ \hline
$\mu _{chem}^{I}$ [MeV] & \multicolumn{1}{|r|}{-14} & -11 & -13 \\
\hline\hline
$\varepsilon _{B}$ [GeV/fm$^{3}$] & 0.48 & 0.14 & 0.42 \\ \hline
$\varepsilon _{M}$ [GeV/fm$^{3}$] & 0.12 & 0.05 & 0.14 \\ \hline
$\varepsilon $ [GeV/fm$^{3}$] & 0.60 & 0.19 & 0.56 \\ \hline
$P$ [GeV/fm$^{3}$] & 0.07 & 0.02 & 0.07 \\ \hline\hline
$\rho _{B}$ [1/fm$^{3}$] & 0.34 & 0.11 & 0.29 \\ \hline
$s$ [1/fm$^{3}$] & 3.4 & 1.3 & 3.3 \\ \hline
$n_{B}$ [1/fm$^{3}$] & 0.34 & 0.11 & 0.29 \\ \hline
$n_{M}$ [1/fm$^{3}$] & 0.17 & 0.09 & 0.19 \\ \hline\hline
$r_{B}$ [GeV] & 1.4 & 1.3 & 1.4 \\ \hline
$r_{M}$ [GeV] & 0.7 & 0.6 & 0.7 \\ \hline
$r$ [GeV] & 1.2 & 1.0 & 1.1 \\ \hline
\end{tabular}
\end{center}
\caption{{\small Thermodynamic parameters at the chemical freeze-out, as
inferred from the analysis of the particle ratios in the Si+Au(Pb) collisions
at AGS. The results correspond to the Table \ref{SiAuRatio}, where the fitted
ratios are shown. }}
\label{SiAu}
\end{table}

The large values of $\chi^2/n$, shown in Table \ref{SiAuRatio},
indicate that thermal description of the particle yields does not work
well for reactions with the silicon beams. Even the ratio of pions to
nucleons does not come out accurately, giving quite large contribution
to $\chi^2/n$. In the thermal analysis of Ref. \cite{PBM-AGS} the
isospin chemical potential is zero, so the ratios such as
$\pi^+/p\,\,$ or $\pi^-/p$ are necessarily the same. The experimental
estimate of Ref.  \cite{PBM-AGS} for any of these ratios is $(2/3)\,
\pi /(p+n)= (2/3) \, 1.05 \approx 0.7$. This result was updated in
Ref. \cite{CleyEllSa}, where the collisions with Au targets are
included only, and the following values are used: $\pi^+/p=0.8$ and
$\pi^-/p=1.0$.  The approach of Ref. \cite{CleyEllSa} has the non-zero
isospin chemical potential, so differences in the abundances of the
states belonging to the same isospin multiplets can be easily
included.  In addition, the authors of Ref. \cite{CleyEllSa} decided
to apply the thermal model to the particle species which are most
abundant.  Consequently, their fit includes only five ratios:
$\pi^+/p, \, K^+/\pi^+, \, K^+/K^-, \, \Lambda/p$ and
$K^-/\pi^-$. Note, that the ratio $\pi^-/p$ is not independent and it
is not included in the calculation.

The results of the thermal fit of Ref. \cite{CleyEllSa} are presented
in Table \ref{CEST5} in the last column denoted by CEST. The
experimental data in the second column are taken also from
Ref. \cite{CleyEllSa}. We have run our code and obtained the results
shown in the third column denoted by TM. First of all, we see that the
five ratios included in the analysis are quite well reproduced in the
thermal model ($\chi^2/n \approx 0.5$).  Our calculation fully
confirms the result of Ref. \cite{CleyEllSa} which gives: $T_{chem} =
110$ MeV and $\mu^B_{chem}=540$ MeV. The four remaining ratios of less
abundant hadron species are treated in this case as an output of the
thermal model. They are shown in the four bottom raws of Table
\ref{CEST5}. Naturally, the agreement of the model in this case is not
as good as for the five ``input'' ratios.

\begin{table}
\begin{center}
\begin{tabular}{|l|l|l|l|}
\hline
{\bf Si+Au} & experiment & TM & CEST \\ \hline\hline
$T_{chem}$ [MeV] &  & \multicolumn{1}{r|}{{\bf 108}$\pm {\bf 5}$} &
\multicolumn{1}{r|}{{\bf 110}$\pm{\bf 5}$} \\ \hline
$\mu _{chem}^{B}$ [MeV] &  & \multicolumn{1}{r|}{{\bf 540}$\pm {\bf 10}$} &
\multicolumn{1}{r|}{{\bf 540}$\pm{\bf 20}$} \\ \hline $\mu _{chem}^{S}$ [MeV] &
& \multicolumn{1}{r|}{93} &  \\ \hline $\mu _{chem}^{I}$ [MeV] &  &
\multicolumn{1}{r|}{-9} &  \\ \hline $\chi ^{2}/n \,(n=5)$ &  &
\multicolumn{1}{r|}{{\bf 0.5}} & \multicolumn{1}{r|}{{\bf 0.6}} \\ \hline\hline
$\pi^{+}/p$ & 0.80$\pm0.08$ & 0.84 & 0.87 \\ \hline $K^{+}/\pi^{+}$ &
0.19$\pm0.02$ & 0.20 & 0.21 \\ \hline $K^{+}/K^{-}$ & 4.40$\pm0.40$ & 4.50 &
4.51 \\ \hline $\Lambda/p$ & 0.20$\pm 0.04$ & 0.14 & 0.16 \\ \hline
$K^{-}/\pi^{-}$ & 0.035$\pm0.005$ & 0.035 & 0.038 \\ \hline \hline
$\Xi^{-}/\Lambda$ & (1.2$\pm0.2$)$\cdot 10^{-1}$ & 5.2$\cdot 10^{-2}$ &
4.9$\cdot 10^{-2}$ \\ \hline $\phi/\pi^{+}$ & (4.5$\pm1.2$)$\cdot 10^{-3}$ &
4.2$\cdot 10^{-3}$ & 4.6$\cdot 10^{-3}$ \\ \hline $\bar{p}/p$ &
(4.5$\pm0.4$)$\cdot 10^{-4}$ & 7.2$\cdot 10^{-5}$ & 7.2$\cdot 10^{-5}$ \\ \hline
$\overline{\Lambda}/\Lambda$ & (2.0$\pm0.8$)$\cdot 10^{-3}$ & 3.2$\cdot 10^{-4}$ &
3.4$\cdot 10^{-4}$ \\ \hline \end{tabular}
\end{center}
\caption{{\small Thermal-model analysis of the particle ratios for Si
+ Au collisions. The experimental data are taken from the paper by
Cleymans et al., see Ref. \cite{CleyEllSa}. The results of
Ref. \cite{CleyEllSa} are given in the last column (CEST) and compared
to our calculation (TM). In the calculation of $\chi^2$ only five
ratios of the most abundant hadron species are included:
$\pi^+/p, \, K^+/\pi^+, \, K^+/K^-, \, \Lambda/p$ and
$K^-/\pi^-$. The remaining
four ratios (the bottom four raws) are treated here as an output of
the thermal model. In our calculation the weak decays were included
again at the 50\% level.  }}
\label{CEST5}
\end{table}

In Table \ref{CEST9} we show the result of the thermal analysis
obtained in the case when all nine ratios are included in the
calculation of $\chi^2$. Our calculation gives $T_{chem} = 127$ MeV
and $\mu^B_{chem}=533$ MeV. As expected, the quality of the fit,
$\chi^2/n \approx 3$, is worse than that obtained in the previous
case with only five ratios included. On the other hand, it is better
than the quality of the fits presented in Table \ref{SiAuRatio}
for Si + Au(Pb) collisions. The better agreement is caused by the
use of the updated ratios of pions to nucleons in Ref. \cite{CleyEllSa}, 
which are more consistent with the thermal picture.

\begin{table}
\begin{center}
\begin{tabular}{|l|l|l|l|}
\hline
{\bf Si+Au} & experiment & TM & CEST \\ \hline\hline
$T_{chem}$ [MeV] &  & \multicolumn{1}{r|}{{\bf 127}$\pm {\bf 3}$} &
\multicolumn{1}{r|}{{\bf 110}$\pm{\bf 5}$} \\ \hline
$\mu _{chem}^{B}$ [MeV] &  & \multicolumn{1}{r|}{{\bf 533}$\pm {\bf 13}$} &
\multicolumn{1}{r|}{{\bf 540}$\pm{\bf 20}$} \\ \hline $\mu _{chem}^{S}$ [MeV] &
& \multicolumn{1}{r|}{116} &  \\ \hline $\mu _{chem}^{I}$ [MeV] &  &
\multicolumn{1}{r|}{-11} &  \\ \hline $\chi ^{2}/n \,(n=9)$ &  &
\multicolumn{1}{r|}{{\bf 3.1}} & \multicolumn{1}{r|}{{\bf 12.2}} \\ \hline\hline
$\pi^{+}/p$ & 0.80$\pm0.08$ & 0.87 & 0.87 \\ \hline $K^{+}/\pi^{+}$ &
0.19$\pm0.02$ & 0.23 & 0.21 \\ \hline $K^{+}/K^{-}$ & 4.40$\pm0.40$ & 4.42 &
4.51 \\ \hline $\Lambda/p$ & 0.20$\pm0.04$ & 0.17 & 0.16 \\ \hline
$K^{-}/\pi^{-}$ & 0.035$\pm0.005$ & 0.042 & 0.038 \\ \hline $\Xi^{-}/\Lambda$ &
(1.2$\pm0.2$)$\cdot 10^{-1}$ & 5.8$\cdot 10^{-2}$ & 4.9$\cdot 10^{-2}$ \\ \hline
$\phi/\pi^{+}$ & (4.5$\pm1.2$)$\cdot 10^{-3}$ & 8.2$\cdot 10^{-3}$ & 4.6$\cdot
10^{-3}$ \\ \hline $\bar{p}/p$ & (4.5$\pm0.4$)$\cdot 10^{-4}$ & 4.1$\cdot
10^{-4}$ & 7.2$\cdot 10^{-5}$ \\ \hline $\overline{\Lambda}/\Lambda$ &
(2.0$\pm0.8$)$\cdot 10^{-3}$ & 1.9$\cdot 10^{-3}$ & 3.4$\cdot 10^{-4}$ \\ \hline
\end{tabular}
\end{center}
\caption{{\small Thermal-model analysis of the particle ratios for Si
+ Au collisions. The experimental data are taken from
\cite{CleyEllSa}. In this case, our calculation (TM) includes all 9
ratios in the construction of $\chi^2$. The last column repeats the
results of Cleymans et al., see Table \ref{CEST5}, which were obtained
by fitting only the first 5  ratios. The value of $\chi^2/n$ is
calculated in both cases (TM and CEST) for all 9 ratios.  }}
\label{CEST9}
\end{table}

\begin{table}
\begin{center}
\begin{tabular}{|l|l|l||l|l||l|l|}
\hline
{\bf S +\ Au(W,Pb)} & experiment & TM & TM 1 & BM 1 & TM 2 & BM 2 \\ \hline\hline
$T_{chem}$ [MeV] &  & \multicolumn{1}{|r||}{{\bf 179}$\pm{\bf 2}$} &
\multicolumn{1}{|r|}{160} & \multicolumn{1}{|r||}{{\bf 160}} &
\multicolumn{1}{|r|}{ 170} & \multicolumn{1}{|r|}{{\bf 170}} \\ \hline
$\mu _{chem}^{B}$ [MeV] &  & \multicolumn{1}{|r||}{{\bf 199}$\pm{\bf 4}$} &
\multicolumn{1}{|r|}{170} & \multicolumn{1}{|r||}{{\bf 170}} &
\multicolumn{1}{|r|}{ 180} & \multicolumn{1}{|r|}{{\bf 180}} \\ \hline
$\mu _{chem}^{S}$ [MeV] &  & \multicolumn{1}{|r||}{55} & \multicolumn{1}{|r|}{
37} & \multicolumn{1}{|r||}{38} & \multicolumn{1}{|r|}{45} &
\multicolumn{1}{|r|}{47} \\ \hline
$\mu _{chem}^{I}$ [MeV] &  & \multicolumn{1}{|r||}{-7} & \multicolumn{1}{|r|}{
-4} & \multicolumn{1}{|r||}{0} & \multicolumn{1}{|r|}{-5} &
\multicolumn{1}{|r|}{0} \\ \hline
$\chi ^{2}/n$ &  & \multicolumn{1}{|r||}{{\bf 4.2}} & \multicolumn{1}{|r|}{11.3}
& \multicolumn{1}{|r||}{{\bf 7.7}} & \multicolumn{1}{|r|}{5.6} &
\multicolumn{1}{|r|}{{\bf 6.6}} \\ \hline\hline
$K_{s\ b}^{0}/\Lambda _{d}$ & 0.88(10) & 1.29 & 1.68 & 1.57 & 1.46 & 1.36 \\
\hline
$K_{s\ b}^{0}/\overline{\Lambda}_{d}$ & 4.6(10) & 5.4 & 7.8 & 7.3 & 6.15 & 5.7 \\
\hline
$\overline{\Lambda }_{c}/\Lambda _{c}$ & 0.20(1) & 0.21 & 0.19 & 0.20 & 0.21
& 0.23 \\ \hline
$\overline{\Lambda }_{d}/\Lambda _{d}$ & 0.19(4) & 0.24 & 0.21 & 0.22 & 0.24
& 0.24 \\ \hline
$\Xi _{c}^{-}/\Lambda _{c}$ & 0.095(6) & 0.112 & 0.12 & 0.12 & 0.16 & 0.12
\\ \hline
$\Xi _{c}^{+}/\overline{\Lambda }_{c}$ & 0.21(2) & 0.20 & 0.18 & 0.20 & 0.19
& 0.21 \\ \hline\hline
$p/\pi ^{+}$ & 0.18(3) & 0.29 & 0.22 & 0.17 & 0.25 & 0.19 \\ \hline
$(h^{+}-h^{-})/h^{-}$ & 0.15(1) & 0.19 & 0.14 & 0.18 & 0.16 & 0.21 \\ \hline
$\left( p-\bar{p}\right) /h^{-}$ & 0.15(2) & 0.20 & 0.15 & 0.13 & 0.17 & 0.14
\\ \hline
$\left( h^{+}-h^{-}\right) /\left( h^{+}+h^{-}\right) $ & 0.088(7) & 0.085 & 
0.067 & 0.084 & 0.075 & 0.094 \\ \hline
$\bar{p}/p$ & 0.12(2) & 0.14 & 0.14 & 0.13 & 0.15 & 0.14 \\ \hline
$\bar{p}/\pi ^{-}$ & 0.024(9) & 0.037 & 0.029 & 0.022 & 0.035 & 0.027 \\ 
\hline
$\eta /\pi ^{0}$ & 0.15(2) & 0.11 & 0.12 & 0.12 & 0.11 & 0.12 \\ \hline
$\phi /\left( \rho +\omega \right) $ & 0.080(20) & 0.055 & 0.053 & 0.11 & 
0.055 & 0.12 \\ \hline
$(K^{+}+K^{-})/(2K_{s}^{0})$ & 1.07(3) & 1.03 & 1.03 & 1.05 & 1.03 & 1.06 \\ 
\hline
$K^{+}/K^{-}$ & 1.67(15) & 1.58 & 1.44 & 1.46 & 1.49 & 1.53 \\ \hline
$K_{s}^{0}/\Lambda $ & 1.4(1) & 1.6 & 2.12 & 1.74 & 1.84 & 1.50 \\ \hline
$K_{s}^{0}/\overline{\Lambda}$ & 6.4(4) & 7.1 & 10.3 & 8.5 & 8.1 & 6.6 \\ \hline
$\Lambda /(p-\bar{p})$ & 0.45(4) & 0.43 & 0.44 & 0.67 & 0.45 & 0.69 \\ \hline
$\overline{\Lambda}/\bar{p}$ & 0.80(30) & 0.62 & 0.55 & 0.38 & 0.59 & 0.41 \\ 
\hline
$\overline{\Lambda}/\Lambda $ & 0.207(12) & 0.226 & 0.206 & 0.20 & 0.226 & 0.23
\\ \hline
$\Xi ^{-}/\Lambda $ & 0.066(13) & 0.114 & 0.120 & 0.12 & 0.118 & 0.12 \\ 
\hline
$\Xi ^{+}/\overline{\Lambda}$ & 0.127(22) & 0.185 & 0.172 & 0.20 & 0.178 & 0.21
\\ \hline
$\Xi ^{+}/\Xi ^{-}$  \, {\small (WA85)} & 0.45(5) & 0.37 & 0.30 & 0.31 & 0.34 & 0.36 \\ \hline
$\Xi ^{+}/\Xi ^{-}$  \, {\small (NA36)} & 0.276(108) & 0.37 & 0.30 & 0.31 & 0.34 & 0.36 \\ \hline
$\frac{\left( \Omega ^{+}+\Omega ^{-}\right) }{\left( \Xi ^{+}+\Xi
^{-}\right) }$ & 0.8(4) & 0.2 & 0.2 & 0.17 & 0.18 & 0.19 \\ \hline
\end{tabular}
\end{center}
\caption{{\small Thermal analysis of the particle ratios measured in
S + Au(W,Pb) reactions. The experimental ratios are taken from
Ref. \cite{PBM-SPS}. The optimal values of the thermodynamic
parameters found in our approach are presented in the third column
(TM).  The results of Ref. \cite{PBM-SPS} are presented in the fifth
and seventh column (denoted by BM1 and BM2, respectively). For
comparison the results of our model for the same input values of
$\,T_{chem}$ and $\mu^B_{chem}$ are shown in the fourth and sixth
column (TM1 and TM2, respectively). Except for the first six measured
ratios (where the weak decays have been disentangled), we include the
weak decays at the level of 50\%. }}
\label{SAuRatio}
\end{table}

\section{S +\ Au(W,Pb)\ Collisions at CERN\ SPS}

The thermal analysis of the S + Au(W,Pb) collisions was performed by
Braun-Munzinger, Stachel, Wessels, and Xu in Ref. \cite{PBM-SPS}.  In
Table \ref{SAuRatio} we show their results (BM1, BM2) together with
our analysis based on the assumption of the full chemical and thermal
equilibrium (TM, TM1, TM2). In our approach we have again neglected
the finite-size and the excluded-volume corrections. The presentation
of the results in Table \ref{SAuRatio} is analogous to that given in
Table \ref{SiAuRatio}. We see again that the values of $\chi^2/n$ are
quite large, indicating that the thermal model describes the measured
ratios only in a qualitative way. 

In Table \ref{SAu} we give the
values of the thermodynamic parameters at the freeze-out. They
exhibit similar features to those found in the case of the Si + Au
collisions discussed in the previous Section. In particular,
we observe the classical equation of state, quite large energy
density and baryon number density, different Cleymans-Redlich ratios
for mesons and baryons.

\begin{table}
\begin{center}
\begin{tabular}{|l|l|l|l|}
\hline
{\bf S +\ Au(W,Pb)} & TM & TM 1 & TM 2 \\ \hline \hline
$T_{chem}$ [MeV] & \multicolumn{1}{|r|}{179$\pm 2$} & \multicolumn{1}{|r|}{
160} & 170 \\ \hline
$\mu _{chem}^{B}$ [MeV] & \multicolumn{1}{|r|}{199$\pm 4$} & 
\multicolumn{1}{|r|}{170} & 180 \\ \hline
$\mu _{chem}^{S}$ [MeV] & \multicolumn{1}{|r|}{55} & \multicolumn{1}{|r|}{37}
& 45 \\ \hline
$\mu _{chem}^{I}$ [MeV] & \multicolumn{1}{|r|}{-7} & \multicolumn{1}{|r|}{-4}
& -5 \\ \hline\hline
$\varepsilon _{B}$ [GeV/fm$^{3}$] & 0.44 & 0.15 & 0.26 \\ \hline
$\varepsilon _{M}$ [GeV/fm$^{3}$] & 0.64 & 0.30 & 0.45 \\ \hline
$\varepsilon $ [GeV/fm$^{3}$] & 1.1 & 0.45 & 0.71 \\ \hline
$P$ [GeV/fm$^{3}$] & 0.16 & 0.07 & 0.11 \\ \hline\hline
$\rho _{B}$ [1/fm$^{3}$] & 0.20 & 0.07 & 0.12 \\ \hline
$s$ [1/fm$^{3}$] & 6.7 & 3.1 & 4.7 \\ \hline
$n_{B}$ [1/fm$^{3}$] & 0.27 & 0.09 & 0.16 \\ \hline
$n_{M}$ [1/fm$^{3}$] & 0.65 & 0.35 & 0.49 \\ \hline\hline
$r_{B}$ [GeV] & 1.6 & 1.6 & 1.6 \\ \hline
$r_{M}$ [GeV] & 1.0 & 0.8 & 0.9 \\ \hline
$r$ [GeV] & 1.2 & 1.0 & 1.1 \\ \hline
\end{tabular}
\end{center}
\caption{{\small
 Thermodynamic parameters at the chemical freeze-out, as
inferred from the analysis of the particle ratios in the S+Au(W,Pb) collisions
at SPS. The results correspond to the cases described in Table \ref{SAuRatio}.
}}
\label{SAu}
\end{table}

\section{Concluding Remarks}

Thermal analysis of the particle ratios measured in the collisions of
relatively lighter nuclei, such as Si or S, leads only to a
qualitative agreement with the data.  This may indicate that the
produced systems are not in the full chemical equilibrium, or they are
too small for the thermodynamic concepts to be applicable.  It is also
possible that the errors of the experimentally determined ratios are
underestimated (the true systematic errors may be larger, note the
discrepancy seen in the measurement of the $\Xi^+/\Xi^-$ ratio done by
different groups).  Our calculations based on the $\chi^2$ method
agree well with the results of Ref. \cite{CleyEllSa}, where only a
limited number of the particle ratios was studied. On the other hand,
we observe appreciable differences between our fits and the global
parametrization of the data proposed in Refs. \cite{PBM-AGS,PBM-SPS}.

\newpage
\thispagestyle{empty}
$\mbox{}$

\chapter{Pb + Pb Collisions at CERN SPS}

Lead on lead collisions at 158 GeV per nucleon were the first really
{\it heavy}\hspace{0.075cm}-ion collisions at fully relativistic energies. In this
case we dealt, for the first time, with large volumes and life-times
of the reaction region. Simple estimates based on the Bjorken
hydrodynamic model \cite{Bjorken} indicate that matter with the energy
density of a few GeV/fm$^3$ was created at the early stage of the
central collisions. Such energy density exceeds the critical energy
density for the phase transition from a hadron gas to a quark-gluon
plasma \cite{Karsch}, so the transient existence of the plasma could occur in
these reactions \cite{press}.

In the first part of this Chapter we show the results of the thermal
analysis of the particle ratios, which is based on our own
implementation of the thermal model. Subsequently, we discuss the
impact of the possible in-medium modifications of hadron masses and
widths on the thermal fits.

\section{Thermal Analysis of Particle Ratios}
\label{tmrat}

\begin{table}
\begin{center}
\begin{tabular}{|l|l|l|l|l|}
\hline
{\bf Pb+Pb} & experiment & max. feeding & no feeding & 50\% feeding \\ \hline\hline
$T_{chem}$ [MeV] &  & \multicolumn{1}{|r|}{{\bf 164}$\pm{\bf 3}$} &
\multicolumn{1}{|r|}{{\bf 170}$\pm{\bf 3}$} & \multicolumn{1}{|r|}{{\bf 168}$\pm
{\bf 3}$} \\ \hline $\mu _{chem}^{B}$ [MeV] &  &
\multicolumn{1}{|r|}{{\bf 234}$\pm{\bf 7}$} & \multicolumn{1}{|r|}{{\bf 250}$\pm
{\bf 8}$} & \multicolumn{1}{|r|}{{\bf 244}$\pm{\bf 8}$} \\ \hline $\mu
_{chem}^{S}$ [MeV] &  & \multicolumn{1}{|r|}{56} & \multicolumn{1}{|r|}{ 65} &
\multicolumn{1}{|r|}{62} \\ \hline $\mu _{chem}^{I}$ [MeV] &  &
\multicolumn{1}{|r|}{-8} & \multicolumn{1}{|r|}{ -9} & \multicolumn{1}{|r|}{-9}
\\ \hline $\chi ^{2}/n$ &  & \multicolumn{1}{|r|}{{\bf 1.4}} &
\multicolumn{1}{|r|}{{\bf 1.8}} & \multicolumn{1}{|r|}{{\bf 0.9}} \\
\hline\hline $\left( p_{a}-\overline{p}_{a}\right) /h_{a}^{-}$ & 0.228(29) &
0.209 & 0.233 & 0.224 \\ \hline
$\overline{p}_{a}/p_{a}$ & 0.055(10) & 0.062 & 0.058 & 0.060 \\ \hline
$\overline{p}_{b}/p_{b}$ & 0.085(8) & 0.083 & 0.080 & 0.081 \\ \hline
$\overline{\Lambda }_{c}/\Lambda _{c}$ & 0.131(17) & 0.118 & 0.118 & 0.118
\\ \hline
$\Xi _{c}^{-}/\Lambda _{c}$ & 0.110(10) & 0.107 & 0.104 & 0.105 \\ \hline
$\Xi _{c}^{+}/\overline{\Lambda }_{c}$ & 0.206(40) & 0.202 & 0.210 & 0.207
\\ \hline\hline
$\pi ^{-}/\pi ^{+}$ & 1.1(1) & 1.2 & 1.1 & 1.1 \\ \hline
$\eta /\pi ^{0}$ & 0.081(13) & 0.099 & 0.121 & 0.109 \\ \hline
$K_{S\ }^{0}/\pi ^{-}$ & 0.125(19) & 0.149 & 0.172 & 0.159 \\ \hline
$K_{S\ }^{0}/h^{-}$ & 0.123(20) & 0.131 & 0.149 & 0.138 \\ \hline
$\Lambda /h^{-}$ & 0.077(11) & 0.102 & 0.076 & 0.091 \\ \hline
$\Lambda /K_{S}^{0}$ & 0.63(8) & 0.78 & 0.51 & 0.66 \\ \hline
$K^{+}/K^{-}$ \,\, {\small (NA44)} & 1.85(9) & 1.68 & 1.78 & 1.74 \\ \hline
$K^{+}/K^{-}$ \,\, {\small (NA49)} & 1.8(1) & 1.68 & 1.78 & 1.74 \\ \hline
$\Xi ^{+}/\overline{\Lambda }$ & 0.188(39) & 0.142 & 0.291 & 0.191 \\ \hline
$\frac{\left( \Xi ^{+}+\Xi ^{-}\right) }{\left( \Lambda +\overline{\Lambda }%
\right) }$ & 0.13(3) & 0.09 & 0.16 & 0.12 \\ \hline
$\Xi ^{+}/\Xi ^{-}$ \,\,\,\,\, {\small (NA49)} & 0.232(33) & 0.223 & 0.237 & 0.232 \\ \hline
$\Xi ^{+}/\Xi ^{-}$ \,\,\,\,\, {\small (WA97)} & 0.247(43) & 0.223 & 0.237 & 0.232 \\ \hline
$\Omega ^{+}/\Omega ^{-}$ & 0.383(81) & 0.448 & 0.522 & 0.493 \\ \hline
$\Omega ^{-}/\Xi ^{-}$ & 0.219(45) & 0.133 & 0.136 & 0.135 \\ \hline
\end{tabular}
\end{center}
\caption{{\small Thermal-model results for the particle ratios measured
in Pb+Pb collisions at CERN SPS. The experimental data are used in the form
compiled in Ref. \cite{PBMHS}.
The contribution from the weak decays
is treated in three different ways: i){\it maximal feeding} (third column)
means that the contributions from all weak decays are included in the 
multiplicities of the produced hadrons, ii) {\it no feeding} (fourth column)
means that the weak decays have been reconstructed, and the contributions 
from the weak decays are not included, and iii) {\it 50\% feeding}
means that 50\% of the contribution from the weak decays is included in
the final multiplicities. }}
\label{PbPbRatio}
\end{table}

With the same implementation of the thermal model as that used in the
previous Chapter to describe the collisions of lighter nuclei, we have
studied the particle ratios measured for Pb + Pb collisions at 158 GeV
per nucleon. We have used the data in the form compiled by
Braun-Munzinger, Heppe, and Stachel in Ref. \cite{PBMHS}. The authors
of Ref. \cite{PBMHS} argued that the particle ratios can be well
described in the framework of the thermal model. Our calculations
support this point of view -- our fits to the particle ratios are
shown in Table \ref{PbPbRatio}. To check the validity of our results,
we did our calculations in three different ways, treating differently
contributions from the weak decays. The three methods were introduced
in the end of Section \ref{dor} (see also Fig. \ref{weakd} where a
scheme of the weak decays is presented). The first 6 ratios listed in
Table \ref{PbPbRatio} were measured in such a way that the feedback
from the weak decays was well established. Consequently, these ratios
enter in the same way the three versions of our calculation. The
remaining 14 ratios are treated differently in each case. For example,
the $\Lambda /h^{-}$ ratio equals: $\Lambda _{d}/h_{b}^{-}$ in the
first version (maximal feeding, shown in the third column), $\Lambda
_{a}/h_{a}^{-}$ in the second version (no feeding, the fourth column),
and $\left( \Lambda _{a}+\Lambda _{d}\right) /\left(
h_{a}^{-}+h_{b}^{-}\right) $ in the third version (averaged feeding,
the fifth column). Other ratios are treated in the similar way. In the
second column the experimental data are shown according to
Ref. \cite{PBMHS}.  We observe that different treatment of the
weak-decay contributions leads to slightly different values of
$T_{chem}, \mu _{chem}^{B},\mu_{chem}^{S}$ and $\mu _{chem}^{I}$, but
the three cases are consistent with each other within errors (compare
the first two raws of Table \ref{PbPbRatio}).

\begin{figure}[h]
\epsfysize=9cm \par
\begin{center}
\mbox{\epsfbox{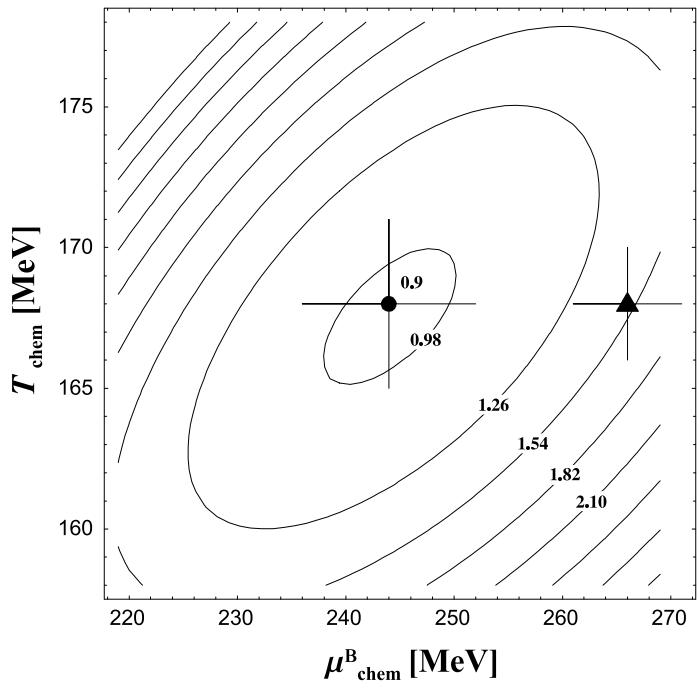}}
\end{center}
\caption{{\small The contour plot of $\chi^2/n$ for the ratios measured
in Pb+Pb collisions at CERN SPS. The numbers at the contours give our
values of $\chi^2/n$. Our optimal values of $T_{chem}$ and $\mu^B_{chem}$
are denoted by the black circle. The optimal values of Ref. \cite{PBMHS}
are denoted by the triangle. The crosses mark the errors. }}
\label{chi2contpb}
\end{figure}

The best fit is obtained when the feeding from the weak decays is
averaged, which corresponds most likely to the experimental
situation. In this case: $T_{chem}= 168\pm 3$ MeV and $\mu
_{chem}^{B}= 244 \pm 8$ MeV. The authors of Ref. \cite{PBMHS} obtained
$T_{chem}= 168\pm 2$ MeV and $\mu _{chem}^{B}= 266 \pm 5$ MeV. The
agreement of the fitted temperatures is very good, the difference of
the fitted values of the baryon chemical potential may be caused by a
different treatment of many branching ratios which are not well known.
In the most realistic case (50\% feeding) our $\chi ^{2}$ per one
degree of freedom is smaller than one, indicating a good quality of
the thermal fit, see Fig. \ref{chi2contpb} where the contour plot of
$\chi^2/n$ is shown.  The total number of degrees of freedom is 20 in
our case. It is smaller by one from the number of ratios included in
Ref. \cite{PBM-SPS}, since we discard the deuteron measurement. The
latter may be disturbed by the inclusion of nuclear fragments.

\begin{table}
\begin{center}
\begin{tabular}{|l|l|l|l|l|l|}
\hline
{\bf Pb+Pb} & experiment &
\begin{tabular}{l}
quantum \\
statistics
\end{tabular}
&
\begin{tabular}{l}
classical \\
statistics
\end{tabular}
& $m<$ 1.8 GeV & $m\leq m_{\Omega }$ \\ \hline\hline
$T_{chem}$ [MeV] &  & \multicolumn{1}{|r|}{{\bf 168}$\pm{\bf 3}$} &
\multicolumn{1}{|r|}{{\bf 167}$\pm{\bf 3}$} & \multicolumn{1}{|r|}{{\bf 171}$\pm
{\bf 3}$} & \multicolumn{1}{|r|}{{\bf 171}$\pm{\bf 3}$} \\ \hline $\mu
_{chem}^{B}$ [MeV] &  & \multicolumn{1}{|r|}{{\bf 244}$\pm{\bf 8}$} &
\multicolumn{1}{|r|}{{\bf 243}$\pm{\bf 8}$} & \multicolumn{1}{|r|}{{\bf 247}$\pm
{\bf 8}$} & \multicolumn{1}{|r|}{{\bf 248}$\pm{\bf 8}$} \\ \hline $\mu
_{chem}^{S}$ [MeV] &  & \multicolumn{1}{|r|}{62} & \multicolumn{1}{|r|}{ 62} &
\multicolumn{1}{|r|}{62} & \multicolumn{1}{|r|}{61} \\ \hline $\mu _{chem}^{I}$
[MeV] &  & \multicolumn{1}{|r|}{-9} & \multicolumn{1}{|r|}{ -9} &
\multicolumn{1}{|r|}{-9} & \multicolumn{1}{|r|}{-8} \\ \hline $\chi ^{2}/n$ &  &
\multicolumn{1}{|r|}{{\bf 0.9}} & \multicolumn{1}{|r|}{{\bf 1.1}} &
\multicolumn{1}{|r|}{{\bf 0.9}} & \multicolumn{1}{|r|}{{\bf 1.1}} \\
\hline\hline $\left( p_{a}-\overline{p}_{a}\right) /h_{a}^{-}$ & 0.228(29) &
0.224 & 0.228 & 0.228 & 0.212 \\ \hline $\overline{p}_{a}/p_{a}$ & 0.055(10) &
0.060 & 0.060 & 0.061 & 0.060 \\ \hline $\overline{p}_{b}/p_{b}$ & 0.085(8) &
0.081 & 0.081 & 0.082 & 0.083 \\ \hline $\overline{\Lambda }_{c}/\Lambda _{c}$ &
0.131(17) & 0.118 & 0.118 & 0.118 & 0.117 \\ \hline
$\Xi _{c}^{-}/\Lambda _{c}$ & 0.110(10) & 0.105 & 0.105 & 0.115 & 0.123 \\
\hline
$\Xi _{c}^{+}/\overline{\Lambda }_{c}$ & 0.206(40) & 0.207 & 0.207 & 0.229 &
0.240 \\ \hline\hline
$\pi ^{-}/\pi ^{+}$ & 1.1(1) & 1.1 & 1.1 & 1.1 & 1.1 \\ \hline
$\eta /\pi ^{0}$ & 0.081(13) & 0.109 & 0.111 & 0.109 & 0.114 \\ \hline
$K_{S\ }^{0}/\pi ^{-}$ & 0.125(19) & 0.159 & 0.162 & 0.156 & 0.161 \\ \hline
$K_{S\ }^{0}/h^{-}$ & 0.123(20) & 0.138 & 0.140 & 0.136 & 0.140 \\ \hline
$\Lambda /h^{-}$ & 0.077(11) & 0.091 & 0.093 & 0.090 & 0.092 \\ \hline
$\Lambda /K_{S}^{0}$ & 0.63(8) & 0.66 & 0.66 & 0.67 & 0.66 \\ \hline
$K^{+}/K^{-}$ \,\,{\small (NA44)} & 1.85(9) & 1.74 & 1.73 & 1.76 & 1.78 
\\ \hline
$K^{+}/K^{-}$ \,\,{\small (NA49)} & 1.8(1) & 1.74 & 1.73 & 1.76 & 1.78 
\\ \hline
$\Xi ^{+}/\overline{\Lambda }$ & 0.188(39) & 0.191 & 0.192 & 0.208 & 0.217
\\ \hline
$\frac{\left( \Xi ^{+}+\Xi ^{-}\right) }{\left( \Lambda +\overline{\Lambda }%
\right) }$ & 0.13(3) & 0.12 & 0.12 & 0.13 & 0.14 \\ \hline
$\Xi ^{+}/\Xi ^{-}$ \,\,\,\,\,{\small (NA49)} & 0.232(33) 
& 0.232 & 0.233 & 0.231 & 0.226 \\ \hline
$\Xi ^{+}/\Xi ^{-}$ \,\,\,\,\,{\small (WA97)} & 0.247(43) 
& 0.232 & 0.233 & 0.231 & 0.226 \\ \hline
$\Omega ^{+}/\Omega ^{-}$ & 0.383(81) & 0.493 & 0.498 & 0.486 & 0.471 \\
\hline
$\Omega ^{-}/\Xi ^{-}$ & 0.219(45) & 0.135 & 0.134 & 0.140 & 0.141 \\ \hline
\end{tabular}
\end{center}
\caption{{\small Thermal-model results for the particle ratios measured
in Pb+Pb collisions at CERN SPS. The experimental data are taken from
Ref. \cite{PBMHS}. The role of the quantum statistics and the mass cut is
displayed.  }}
\label{PbPbStat}
\end{table}

\section{Sensitivity to Classical Statistics and Limited Mass Spectrum}
\label{sens}

In this Section, we discuss the sensitivity of our results to two
different modifications. The first modification is simply a
replacement of the quantum distribution functions by the classical
distributions (the Bose-Einstein distribution for mesons and the
Fermi-Dirac distribution for baryons are  replaced by the Boltzmann
distribution for all hadrons including pions and kaons). The second
modification is connected with the introduction of a limiting hadron
mass. In this way we can check how important the tail of the mass
distribution of hadrons is for the results of our analysis.

The results obtained with the classical distribution functions and
with the limited hadron mass spectrum are shown in Table
\ref{PbPbStat}. The data are listed in the second column. In the third
column we present our results obtained for exact quantum
statistics. They coincide with the last column of Table
\ref{PbPbRatio}, i.e., we average the feeding from the weak decays
here.  In the fourth column we show the results for the classical
statistics. The fifth and sixth columns show our results obtained with
the limited mass spectrum. In the fifth column we neglect the feedback
from the resonances heavier than 1.8 GeV, whereas in the sixth column
the maximal mass is equal to the mass of $\ \Omega $ ($m_\Omega=$ 1.672
GeV). In each column we show the corresponding values of the optimal
thermodynamic parameters and $\chi ^{2}/n$.

We find that the use of the classical distribution functions leads to
very small changes of our results. The values of $T_{chem}$ and
$\mu^B_{chem}$ change only by 1 MeV, and $\chi^2/n$ increases from 0.9
to 1.1.  The quantum statistics are not important in the thermal
analysis because most of particles are very heavy and they can be
described well by the classical distributions. Even for pions we can
use the classical statistics, since they are produced mainly by the
decays of heavier resonances.  The classical features of the hadronic
fireball at the freeze-out are also reflected in the equation of
state, which has a form $P=(n_M+n_B)T$.  Cutting the mass spectrum at
1.8 GeV changes the values of $T_{chem}$ and $\mu^B_{chem}$ by 3
MeV. The cut at $m_\Omega=$ 1.672 GeV has a similar impact on
$T_{chem}$ and $\mu^B_{chem}$, which remain in agreement (within
errors) with the exact results. We have checked that the cuts at
smaller masses, $\sim 1.6$ GeV, cause already appreciable changes of
$T_{chem}$ and $\mu^B_{chem}$.

\section{Thermodynamic Properties of  Freeze-Out}
\label{thermo}

In this Section we present the complete set of the thermodynamic
parameters characterizing the chemical freeze-out in Pb + Pb
collisions at CERN SPS. We have again done our calculations in three
different ways, treating differently contributions from the weak
decays. As discussed above, in the three considered cases the values
of $T_{chem}$ and $\mu^B_{chem}$ are consistent with each other.  It
is interesting to observe, however, that small differences in
$T_{chem}$ and $\mu^B_{chem}$ may result in quite large changes of
other thermodynamic parameters, obtained as the integrals over the
distribution functions (\ref{fip}). In Table \ref{PbPb} we show the
energy density, pressure, baryon number density, entropy, baryon and
meson densities, and the Cleymans-Redlich ratios.  We find that
different treatment of the weak decays causes that the thermal-model
estimate of the energy density at the chemical freeze-out varies from
0.6 GeV/fm$^3$ to 0.8 GeV/fm$^3$. Similarly, the baryon density at the
freeze-out changes from 0.13 fm$^{-3}$ to 0.19 fm$^{-3}$. Such changes
indicate, that the correct experimental reconstruction of the weak
decays is necessary in order to have more precise information about
the state of matter at the freeze-out.  The last three raws of Table
\ref{PbPb} show the Cleymans-Redlich ratio calculated for baryons
($r_B$), mesons ($r_M$), and all hadrons ($r$).  We observe that $r$
is close to unity in the three considered cases, i.e., it is
insensitive to the way in which the weak decays are treated. We again
find that $r_B$ is much larger than $r_M$.

\begin{table}
\begin{center}
\begin{tabular}{|l|l|l|l|}
\hline
{\bf Pb+Pb} & Max. feeding & No feeding & 50 \% \ feeding \\ \hline \hline
$T_{chem}$ [MeV] & \multicolumn{1}{|r|}{164$\pm 3$} & \multicolumn{1}{|r|}{
170$\pm 3$} & \multicolumn{1}{|r|}{168$\pm 3$} \\ \hline
$\mu _{chem}^{B}$ [MeV] & \multicolumn{1}{|r|}{234$\pm 7$} &
\multicolumn{1}{|r|}{250$\pm 8$} & \multicolumn{1}{|r|}{244$\pm 8$} \\ \hline
$\mu _{chem}^{S}$ [MeV] & \multicolumn{1}{|r|}{56} & \multicolumn{1}{|r|}{65}
& \multicolumn{1}{|r|}{62} \\ \hline
$\mu _{chem}^{I}$ [MeV] & \multicolumn{1}{|r|}{-8} & \multicolumn{1}{|r|}{-9}
& \multicolumn{1}{|r|}{-9} \\ \hline\hline
$\varepsilon _{B}$ [GeV/fm$^{3}$] & 0.24 & 0.35 & 0.31 \\ \hline
$\varepsilon _{M}$ [GeV/fm$^{3}$] & 0.35 & 0.45 & 0.41 \\ \hline
$\varepsilon $ [GeV/fm$^{3}$] & 0.59 & 0.80 & 0.72 \\ \hline
$P$ [GeV/fm$^{3}$] & 0.09 & 0.12 & 0.11 \\ \hline\hline
$\rho _{B}$ [1/fm$^{3}$] & 0.13 & 0.19 & 0.17 \\ \hline
$s$ [1/fm$^{3}$] & 4.0 & 5.1 & 4.7 \\ \hline
$n_{B}$ [1/fm$^{3}$] & 0.15 & 0.22 & 0.19 \\ \hline
$n_{M}$ [1/fm$^{3}$] & 0.40 & 0.49 & 0.46 \\ \hline\hline
$r_{B}$ [GeV] & 1.6 & 1.6 & 1.6 \\ \hline
$r_{M}$ [GeV] & 0.9 & 0.9 & 0.9 \\ \hline
$r$ [GeV] & 1.1 & 1.1 & 1.1 \\ \hline
\end{tabular}
\end{center}
\caption{{\small Thermodynamic parameters at the chemical freeze-out
in Pb + Pb collisions at CERN SPS. Different treatment
of the weak decays leads to quite large changes in the estimates
of the energy density, baryon number density, and other thermodynamic
quantities. }}
\label{PbPb}
\end{table}

\section{Scaling of Masses and Widths at Freeze-Out}
\label{scala}

Thermal-model fits show that the temperature at the chemical
freeze-out, $T_{chem}$, as well as the baryon density, are large. As
we have just seen in the previous Sections, one typically obtains
$T_{chem}\sim $ 170 MeV, which is close to the expected critical value
for the deconfinement/hadronization phase transition \cite{Karsch}. In
this situation one may expect that hadron properties at the chemical
freeze-out are strongly modified by the presence of the hadronic
environment. Indeed, such modifications are predicted by different
model calculations 
%\cite{hatsuda,klingl,HatsudaLee,BR}
$[28-31]$, which helps to
explain the low-mass dilepton enhancement observed in the CERES
\cite{CERES} and HELIOS \cite{HELIOS} experiments. In this Section we
incorporate possible modifications of hadron masses and widths into
thermal analysis of the particle ratios. We generalize the results of
Refs. \cite{FlorBron,Hirsch} where the problem was studied without
refitting thermodynamic parameters.

Initially,  we include only the mass modifications and calculate
the particle densities from the ideal-gas expression%
\begin{equation}
n_{i}=\frac{g_{i}}{2\pi ^{2}}\int_{0}^{\infty }\frac{p^{2}\ dp}{\exp \left[
\left( E_{i}^{\ast }-\mu _{chem}^{B}B_{i}-\mu _{chem}^{S}S_{i}-\mu
_{chem}^{I}I_{i}\right) /T_{chem}\right] + \epsilon} \, ,
\label{Nistar}
\end{equation}
where
\begin{equation}
E_{i}^{\ast }=\sqrt{p^{2}+\left( m_{i}^{\ast}\right) ^{2}}
\label{Eipstar}
\end{equation}
is the energy. Of course, in standard thermal-model fits Eq. (\ref{Nistar})
is used with the vacuum masses, $m_{i}^{\ast }=m_{i}$. The in-medium masses,
$m_{i}^{\ast }$, may depend on temperature and density in a complicated way.
In order to explore possible different behavior of in-medium masses and, at
the same time, keep simplicity, we do our calculations with the meson and
baryon masses rescaled by the two universal parameters, $x_{M}$ and $x_{B}$,
namely
\begin{equation}
m_{M}^{\ast }=x_{M}\,m_{M},\hspace{0.5cm}m_{B}^{\ast }=x_{B}\,m_{B}.
\label{iksy}
\end{equation}
An exception from this rule are the masses of pseudo-Goldstone bosons
($\pi$, $K$, $\eta$) which we keep constant. This is in agreement with
explicit model calculations incorporating chiral symmetry, {\it e.g.}
\cite{c1,c2}. We note that the use of Eq. (\ref{Nistar}) is valid when
the in-medium hadrons can be regarded as good quasi-particles. A
thermodynamically consistent approach has been constructed so far only
for the lowest multiplets of hadrons \cite{zsch}. At SPS energies,
however, it is crucial to include all hadrons with masses up to (at
least) 1.6 GeV. For such a complicated system, a thermodynamically
consistent approach is not available.  As in the standard approach,
Eq. (\ref{Nistar}) is used to calculate the primordial density of
stable hadrons and resonances at the chemical freeze-out. The final
(observed) multiplicities receive contributions from the primordial
stable hadrons, and from the secondary hadrons produced by decays of
resonances after the freeze-out. Ref. \cite{PDG} is used to determine
the branching ratios, which we keep unchanged. We neglect the
finite-size and excluded volume corrections which do not affect the
particle ratios.

For fixed values of $x_{M}$ and $x_{B}$ we fit the temperature,
$T_{chem}$, and the baryonic chemical potential, $\mu _{chem}^{B}$, by
the minimization the expression
\begin{equation}
\chi ^{2}(x_M,x_B) = \sum\limits_{k=1}^n 
{ \left[ R_{k}^{\,exp}-R_{k}^{\,therm}(x_M,x_B)\right] ^{2}
\over \sigma _{k}^{2}},
\label{chi2x}
\end{equation} 
where $R_{k}^{\,exp}$ is the $k$th measured ratio, $\sigma _{k}$ is
the corresponding error, and $R_{k}^{\,therm}(x_M,x_B)$ is the same
ratio as determined from the thermal model with the modified
masses. Similarly to the standard case with unchanged masses, $m_i^* =
m_i$, the strange chemical potential, $\mu _{chem}^{S},$  and the
isospin chemical potential, $\mu _{chem}^{I}$, are determined from the
requirements that the initial strangeness of the system is zero, and
that the ratio of the baryon number to the electric charge is the same
as in the colliding nuclei, see Eqs. (\ref{s0}) and (\ref{zovera}).

\begin{figure}[h]
\epsfysize=13.0cm
\par
\begin{center}
\mbox{\epsfbox{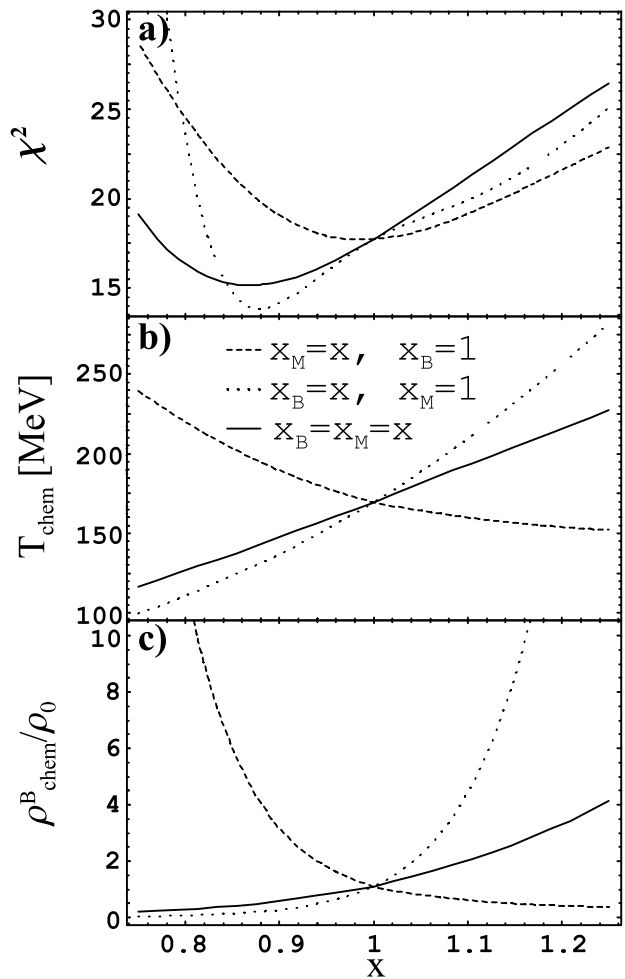}}
\end{center}
\caption{{\small Dependence of $\chi^{2}$, and the fitted values of
the temperature and the baryon density on the scale parameter
$x$. Plot represents Pb + Pb collisions at SPS energies. Solid lines:
all hadron masses (except for Goldstone bosons) are scaled with
$x_{M}=x_{B}=x$. Dashed line: only meson masses are scaled,
$x_{M}=x,x_{B}=1$. Dotted lines: only baryon masses are scaled,
$x_{B}=x,x_{M}=1$. The nuclear saturation density $\rho_{0}$ = 
0.17 fm$^{-3}$.}}
\label{iks}
\end{figure}

In Fig. \ref{iks} we plot our results obtained for the experimental
ratios measured in Pb + Pb collisions at SPS. In the case
$x_{M}=x_{B}=1$ we recover our results discussed in detail in Sections
\ref{tmrat} -- \ref{thermo}.  In Fig. \ref{iks} a) we give our values
of $\chi ^{2}$. One can observe that a small decrease of the meson and
baryon masses, $x_{M}=x_{B}\sim 0.9$, leads to a slightly better fit
with the corresponding smaller values of the temperature and the
baryon density, as shown in Figs. \ref{iks} b) and \ref{iks} c).  It
would be premature, however, to conclude that the masses drop. The
values of $\chi^2$ for the solid line are increased by 25\% compared
to the minimum in the range $0.75 < x < 1.05$, which clearly is the
allowed range.  We thus conclude that moderate dropping of hadron
masses, say by 20\%, does not spoil the quality of thermal fits. On
the contrary, larger dropping or growing of the masses result in a
significant increase of the $\chi^2$ values.

With the modified masses the thermodynamic parameters characterizing
the fits change. For example, if we rescale both meson and baryon
masses (except for Goldstone bosons) in the same way, $x=x_{M}=x_{B}$,
the temperature and the chemical potentials \ are to a very good
approximation also rescaled by $x$. This follows from the fact that we
study a system of equations which is invariant under rescaling of all
quantities with the dimension of energy. If we allowed also for the
changes of the masses of the Goldstone bosons, the thermodynamic
parameters would scale exactly as $T_{chem}(x)=x\ T_{chem}(x=0)$ and
$\mu _{chem}(x)=x\ \mu _{chem}(x=0)$. In this case $\chi ^{2}$ remains
constant, independently of $x$. For fixed values of the
Goldstone-boson masses, the scale invariance is broken, $\chi ^{2}$
varies with $x$, as shown in Fig. \ref{iks} a), and the results are
non-trivial.

To account for finite in-medium widths, $\Gamma _{i}^{\ast }$, of the
resonances we generalize Eq. (\ref{Nistar}) to the formula 
%\cite{BU,DMB,WW,WFN}
[57-60]
\begin{eqnarray}
\hspace{-0.5cm} &&n_{i}=\int\limits_{M_{0}^{2}}^{\infty
}dM^{2}\int\limits_{0}^{\infty }dp\,\,{\frac{1}{\pi N}}{\frac{m_{i}^{\ast
}\Gamma _{i}^{\ast }}{(M^{2}-(m_{i}^{\ast })^{2})^{2}+(m_{i}^{\ast }\Gamma
_{i}^{\ast })^{2}}}  \nonumber \\
\hspace{-0.5cm} &&\times \frac{g_{i}}{2\pi ^{2}}\frac{p^{2}}{\exp \left[
\left( \sqrt{M^{2}+p^{2}}-\mu _{chem}^{B}B_{i}-\mu _{chem}^{S}S_{i}-\mu
_{chem}^{I}I_{i}\right) /T_{chem}\right] \pm 1},  \label{nigamma}
\end{eqnarray}
where $N$ is the normalization of the relativistic Breit-Wigner function,
\begin{equation}
N=
{\frac{1}{2}}+{\frac{1}{\pi }}\arctan \left[ ((m_{i}^{\ast
})^{2}-M_{0}^{2})/(m_{i}^{\ast }\Gamma _{i}^{\ast })\right] \approx 1.
\end{equation} 
The integral over $M^{2}$ is taken to start at the threshold
$M_{0}^{2}$ corresponding to the dominant decay channel. In the limit
$\Gamma _{i}^{\ast }\rightarrow 0$ Eq. (\ref{nigamma}) obviously
reduces to formula (\ref{Nistar}).

\begin{figure}[t ]
\epsfysize=13cm
\par
\begin{center}
\mbox{\epsfbox{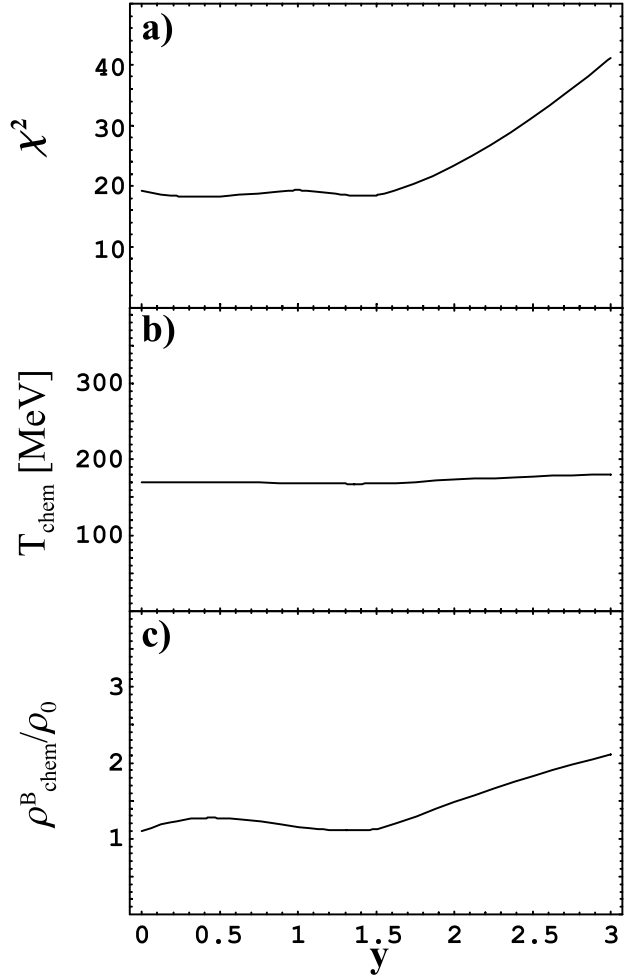}}
\end{center}
\caption{{\small Dependence of $\protect\chi^2$, and the optimal values of the
temperature and the baryon density on the scale parameter for the widths, $y$.}}
\label{yksy}
\end{figure}

In order to analyze the effect of broadening of hadron widths we
introduce the parameter $y$ in such a way that
\begin{equation}
\Gamma _{i}^{\ast }=y\,\Gamma _{i}.
\end{equation}
Here $\Gamma _{i}$ are the vacuum widths, hence the case $y=1$
corresponds to the physical widths as measured in the vacuum, and the
case $y=0$ represents the situation when the widths are neglected (our
previous analysis based on Eq. (\ref{Nistar})). In Fig. \ref{yksy} we
show the results of our fitting procedure. We observe that the
inclusion of the vacuum widths does not change the value of $\chi
^{2}$, and the values of $T_{chem}$ and $\rho _{chem}^{B}$. An
increase of the widths by a factor of 2 has also little effect. Only
for larger modifications of the widths the fit gets worse.

In conclusion we state that the thermal analysis of particle ratios,
measured in Pb + Pb collisions at CERN SPS, allows for moderate
dropping of hadron masses ($\sim$ 20\%).  This does not spoil the
fits, which remain of similar quality as those obtained without
modifications.  Larger dropping of the hadron masses or growing of the
masses are not likely.  Scaling of hadron masses results in
modifications of the thermodynamic parameters for which the fits are
optimal. In particular, lowering of \ all the masses leads to a
smaller values of $T_{chem}$ and $\rho _{chem}^{B}$. This might be a
desired effect, since $T_{chem}\sim $170 MeV is large and may
correspond to quark-gluon plasma rather than to a hadron gas. Our
study of the modifications of the hadron widths shows that they have
small impact on the ratios.

\chapter{Au + Au Collisions at BNL RHIC}

The heavy-ion program at CERN SPS delivered several very interesting
results, but there was no clear discovery of the new physical
phenomena, such as the desired deconfinement phase transition. The
newest heavy-ion machine, the Relativistic Heavy-Ion Collider
constructed at BNL, was designed to accelerate heavy ions at energy
$\sqrt{s}=$ 200 GeV. In the year 2000, during the first run of RHIC,
the maximal energy of $\sqrt{s}=$ 130 GeV was achieved. This energy
exceeds the CERN SPS energy by one order of magnitude
\footnote{The energy of 158 GeV per nucleon in the lab corresponds to
$\sqrt{s}=$ 17 GeV per nucleon pair in the center-of-mass system.},
so the new phenomena were expected to occur. The first new RHIC data
indicate indeed at several new features of the collision process such
as: much higher particle multiplicities, increased production of
antiparticles, and lower baryon number in the central region.
Nevertheless, the overall picture of the collision follows the pattern
established from the studies of heavy-ion collisions at lower energies.
Further systematic analysis of the just incoming data is necessary to extract
more information.

\section{Thermal Analysis of Particle Ratios}

Table \ref{RHICratios} presents our fit to the particle ratios
measured at RHIC during its first run at $\sqrt{s}=$ 130 GeV.  We
stress that exactly the same version of the thermal model has been
used in this fit, as that used in our previous studies of Si + Au, S +
Au, and Pb + Pb reactions. In our present calculation, the identical
ratios measured by different groups are treated separately in the
definition of $\chi^{2}$ (number of points $n=16$). In this way the
measurements done by different groups enter independently, and
converging experimental data have a larger weight in the statistical
analysis. Very similar results are obtained, however, if we first
average the results of different groups to obtain the most likely
value for each considered ratio. This fact shows consistency
of the experimental measurements done by different groups, see the
third column in Table \ref{RHICratios}.

Our optimal value of $T_{chem}$ obtained from the analysis of the
particle ratios equals $165 \pm 7$ MeV.  It is a very interesting fact
that $T_{chem}$ for RHIC agrees well with $T_{chem}=168 \pm 3$ MeV
found in our fit for Pb + Pb collisions at CERN SPS. The growth of the
energy of the colliding nuclei by one order of magnitude does not lead
to creation of a hotter system, only the baryon chemical potential is
significantly lower than that found at CERN SPS. Clearly, the
collisions at RHIC energies are more transparent than the collisions
at CERN energies.  We have also calculated other characteristics of
the freeze-out.  In particular, we find the energy density
$\varepsilon =0.5$ GeV/fm$^{3}$, the pressure $P=$ 0.08 GeV/fm$^{3}$,
and the baryon density $\rho _{B}=$ 0.02 fm$^{-3}$. Here again, one
can observe that the energy density is not higher than that found at
CERN SPS, only the net baryon number tends rapidly to zero. Our
calculation confirms the Cleymans-Redlich conjecture \cite{CleyRed}
that the energy per hadron at the chemical freeze-out is 1 GeV (our
approach yields almost exactly $r =$ 1.0 GeV). We observe, however,
that the average baryon energy is much larger than the average
meson energy: $r_B =$ 1.6 GeV and $r_M =$ 0.9 GeV. We emphasize
that exactly the same $r_B$ and $r_M$ have been extracted from Pb + Pb
collisions at CERN SPS, see Table \ref{PbPb}. Also the values of $r_B$
and $r_M$ extracted from the sulphur collisions, see Table \ref{SAu},
agree well with these two values. Further work is needed to explain
such universal behavior. In addition, in our thermal approach we find
that the ratios $\overline{\Lambda } /\Lambda $ and $\overline{\Xi
}/\Xi $ are practically unaffected by the weak decays, since the
latter contribute in the same way to the abundances of baryons and
antibaryons (note the very small value of our $\mu^B_{chem}$).

%\savebox{\tabRHIC}{\vbox{
\begin{table}
\begin{center}
\begin{tabular}{|l|l|c|}
\hline
{\bf Au+Au} & TM & experiment \\ \hline \hline
$T_{chem}$ [MeV] & \multicolumn{1}{|r|}{{\bf 165}$\pm {\bf 7}$} &  \\ \hline
$\mu _{chem}^{B}$ [MeV] & \multicolumn{1}{|r|}{{\bf 41}$\pm {\bf 5}$} &  \\
\hline $\mu _{chem}^{S}$ [MeV] & \multicolumn{1}{|r|}{9} &  \\ \hline
$\mu _{chem}^{I}$ [MeV] & \multicolumn{1}{|r|}{-1} &  \\ \hline
$\chi ^{2}/n$ & \multicolumn{1}{|r|}{{\bf 0.97}} &  \\ \hline \hline
$\pi ^{-}/\pi ^{+}$ & $1.02$ &
\begin{tabular}{ll}
$1.00\pm 0.02$ \cite{phobos}, & $0.99\pm 0.02$\cite{bearden}
\end{tabular}
\\ \hline
$\overline{p}/\pi ^{-}$ & $0.09$ & $0.08\pm 0.01$ \cite{harris} \\ \hline
$K^{-}/K^{+}$ & $0.92$ &
\begin{tabular}{ll}
$0.88\pm 0.05$ \cite{caines}, & $0.78\pm 0.12$ \cite{ohnishi} \\
$0.91\pm 0.09$ \cite{phobos}, & $0.92\pm 0.06$ \cite{bearden}
\end{tabular}
\\ \hline
$K^{-}/\pi ^{-}$ & $0.16$ & $0.15\pm 0.02$ \cite{caines} \\ \hline
$K_{0}^{\ast }/h^{-}$ & $0.046$ & $0.060\pm 0.012$ \cite{caines,zxu} \\
\hline
$\overline{K_{0}^{\ast }}/h^{-}$ & $0.041$ & $0.058\pm 0.012$ \cite
{caines,zxu} \\ \hline
$\overline{p}/p$ & $0.65$ &
\begin{tabular}{ll}
$0.61\pm 0.07$ \cite{harris}, & $0.54\pm 0.08$ \cite{ohnishi} \\
$0.60\pm 0.07$ \cite{phobos}, & $0.61\pm 0.06$ \cite{bearden}
\end{tabular}
\\ \hline
$\overline{\Lambda }/\Lambda $ & $0.69$ & $0.73\pm 0.03$ \cite{caines} \\
\hline
$\overline{\Xi }/\Xi $ & $0.76$ & $0.82\pm 0.08$ \cite{caines} \\ \hline
\end{tabular}%}}
%\begin{table}[t]
%\vspace{1.0cm}
%\par
%\begin{center}
%\usebox{\tabRHIC}
\end{center}
\caption{{\small Thermal fit of the particle ratios measured at RHIC
at $\sqrt{s}=130$ GeV.}}
\label{RHICratios}
\end{table}

\begin{figure}[h]
\epsfysize=11cm \par
\begin{center}
\mbox{\epsfbox{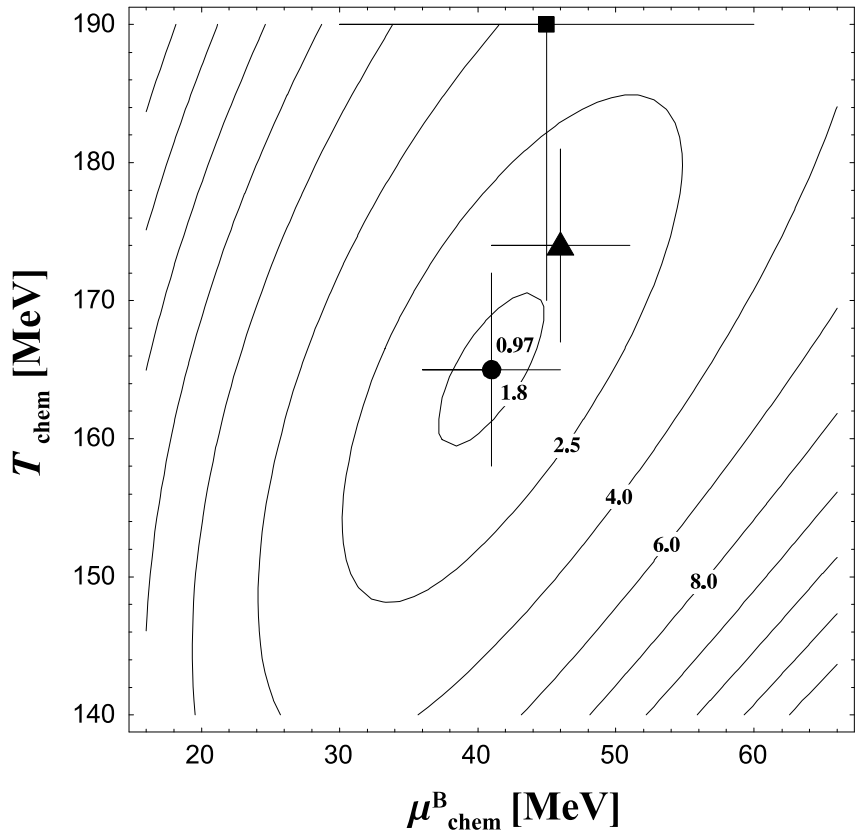}}
\end{center}
\caption{{\small The contour plot of our  $\chi^2/n$. Our result (black
circle), the result of Ref. \cite{nxu} (black square), and the fit of
Ref. \cite{pbmrhic} (black triangle) are all shown with the corresponding
errors. The numbers at the contours indicate the values of $\chi^2/n$.}}
\label{chi2cont}
\end{figure}

In Fig. \ref{chi2cont} we show $\chi^2/n$ as a function of the
temperature, $T_{chem}$, and the baryon chemical potential,
$\mu^B_{chem}$. An interesting feature of the plot is that it shows a
characteristic valley of the optimal values of the thermodynamic
parameters. The shape of this valley indicates that the quality of the
fit does not change much if we moderately increase or decrease both
$T_{chem}$ and $\mu^B_{chem}$. We note that a similar valley can be
seen in Fig. \ref{chi2contpb}, where the values of $\chi^2/n$ were
plotted for the case of Pb + Pb collisions at CERN SPS.  While our work
was nearing completion \cite{FlorBronMich-TherAnal}, a fit by
Braun-Munzinger, Magestro, Redlich and Stachel was presented:
$T_{chem}=174 \pm 7$ MeV and $\mu^B_{chem}=46 \pm 5$ MeV
\cite{pbmrhic}. We note that our $T_{chem}$ is 9 MeV lower than
$T_{chem}$ of Ref. \cite{pbmrhic}, and 25 MeV lower than 190 MeV
obtained in a similar fit by Xu and Kaneta \cite{nxu}. Still, the
results of the three calculations are consistent within errors.  This
fact is displayed in Fig. \ref{chi2cont}, where our result (black
circle), the fit of Ref. \cite{pbmrhic} (black triangle), and the fit
of Ref. \cite{nxu} (black square) are shown all together with the
corresponding errors. Especially, the results of Refs. \cite{pbmrhic}
and \cite{FlorBronMich-TherAnal} are close and line up
along the valley of the optimal parameters.

\begin{figure}[t]
\epsfysize=8cm \par
\begin{center}
\mbox{\epsfbox{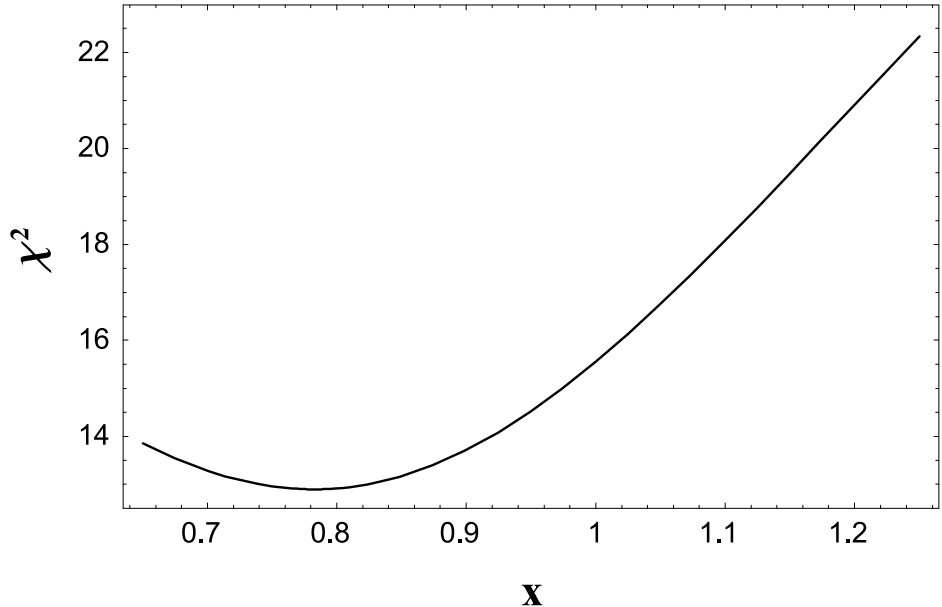}}
\end{center}
\caption{{\small Dependence of $\chi^2$ on the scale parameter $x$.
In this case $x_B=x_M=x$, i.e., the masses of baryons and mesons
are rescaled in the same way (except for the pseudo-Goldstone bosons).  }}
\label{chi2x2}
\end{figure}

Similarly to the case of Pb + Pb collisions at CERN SPS, we have
studied the influence of possible in-medium mass modifications on the
thermal analysis of the ratios measured at RHIC. In Fig.  \ref{chi2x2}
we show the dependence of $\chi^2$ on the scale parameter $x$ (in this
case the mass modifications of baryons and mesons are assumed to be
equal). We observe that the values of $\chi^2$ are slightly lowered
for $x < 1$. This is analogous behavior to that observed in Pb + Pb
collisions studied in Sect.  4.4. Again, the effect is not dramatic
($\chi^2$ decreases by 20\% if $x$ changes from 1 to 0.8), so the
indication for dropping masses is rather weak. On the other hand, we
see that the changes of the masses can be included in the thermal
analysis, yielding the fits of the same quality as that found in the
calculations without any modifications. In the present case, however,
we see that larger dropping of the masses is allowed, as compared to
the CERN-SPS case discussed before.  We have also analyzed Si + Au
collisions at AGS, and S +\ Au collisions at SPS. In these two cases
$\chi ^{2}$ has a flat minimum at $x_{M}\approx x_{B}\approx 1$.

\section{Pressure of Hadron Gas at Freeze-Out}

\begin{figure}[h]
\epsfysize=8.0cm \par
\begin{center}
\mbox{\epsfbox{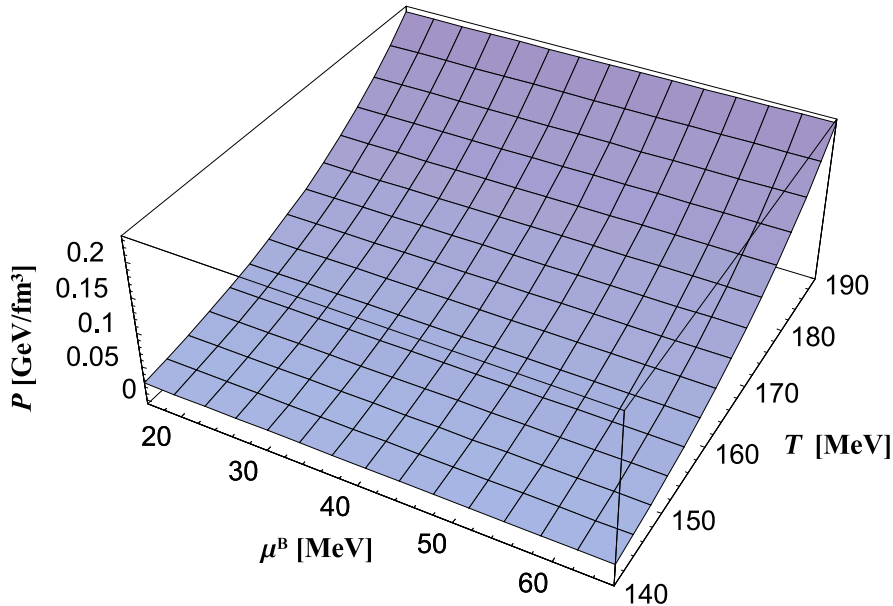}}
\end{center}
\caption{{\small Pressure of a hadron gas which includes all resonances
consisting of up, down, and strange quarks. }}
\label{p}
\end{figure}

In Fig. \ref{p} we plot the pressure of a hadron gas as a function of
the temperature and the baryon chemical potential. We show the
neighborhood of the optimal parameters: $T_{chem}=165$ MeV and
$\mu^B_{chem}=41$ MeV.  The exact result shown in Fig. \ref{p},
obtained as a sum of all partial pressures given by Eq. (\ref{pres}),
can be very well approximated by the formula
\begin{eqnarray}
P_{HG}(T,\mu^B) &=& {1 \over 2 \pi^2} \int\limits_{0}^{M_{\max}}
\rho_M(m)\, m^2 \,  T^2 \, K_2 \left( {m \over T} \right) \, dm \nonumber \\
& & +  {1 \over \pi^2} \int\limits_{0}^{M_{\max}}
\rho_B(m)\,\hbox{cosh}\left( {\mu^B \over T} 
\right) m^2 \, T^2 \, K_2 \left( {m \over T} \right) \, dm \nonumber \\
&=&  {1 \over 2  \pi^2} \int\limits_{0}^{M_{\max}}
\left[ \rho_M(m)+ 2 \,  \hbox{cosh}\left( {\mu^B \over T} 
\right) \rho_B(m) \right]  m^2 \, T^2 \, K_2 \left( {m \over T} \right) \, dm.
\label{phg}
\end{eqnarray}
In Eq. (\ref{phg}) the sum over all hadronic states is replaced by the
convolution of the classical formula for the partial pressure,
Eq. (\ref{claepsp}), with the mass spectrum for baryons and mesons,
Eqs. (\ref{tbh}) and (\ref{tmh}). The maximal mass included in the
integral is taken to be the same as that used in the fits of the
spectrum \cite{brohag2}, $M_{\max}=1.8$ GeV.  We note that the strange
chemical potential as well as the isospin chemical potential have been
neglected in Eq. (\ref{phg}). Our fit shows that these two potential
are very small  at RHIC. For $T=165$ MeV and $\mu^B=41$ MeV, the
pressure calculated from Eq. (\ref{phg}) agrees very well with the
exact result $P=0.08$ GeV/fm$^3$. The very weak dependence of the
pressure on the baryon chemical potential, as displayed in
Fig. \ref{p}, follows also from Eq. (\ref{phg}), since in the considered
region one finds $\hbox{cosh\,}(\mu^B/T) \approx 1$.

\begin{figure}[t]
\epsfysize=10cm
\par
\begin{center}
\mbox{\epsfbox{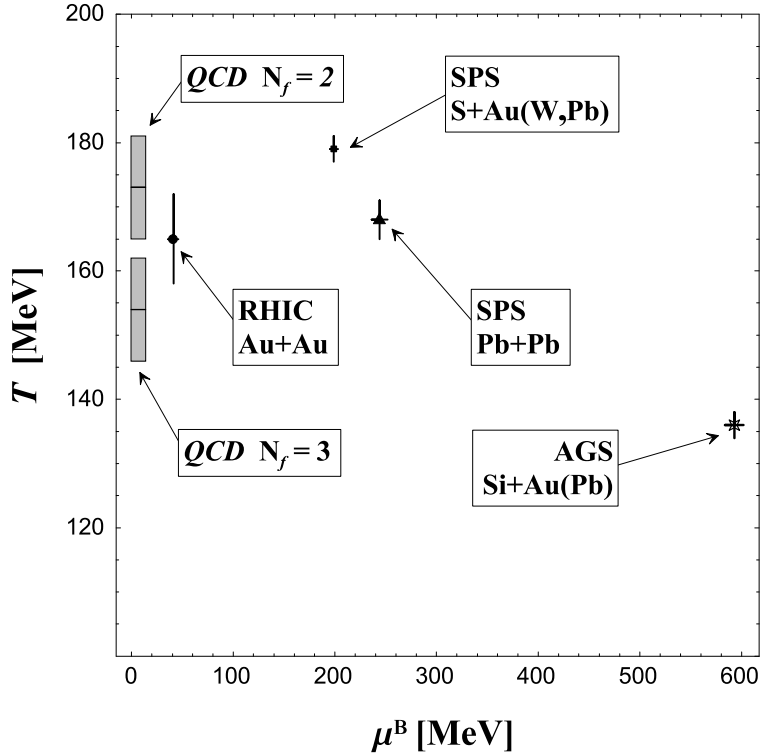}}
\end{center}
\caption{{\small Our optimal thermodynamic parameters.  The fitted
values of the temperature for Pb + Pb collisions at CERN SPS and for
Au + Au collisions at BNL RHIC are consistent with each other, and
close to the critical temperature predicted by the QCD simulations on
a lattice \cite{Karsch}.  The lattice results (grey bands) correspond
to two different versions of the calculations (the number of massless
flavors, $N_f$, is taken to be either 2 or 3). The vertical size of the 
bands denotes the error of the lattice calculations. We show also our
optimal parameters describing the collisions of lighter nuclei. Note
that in this case the quality of the fits is worse, as discussed in
detail in Chapter 3. }}
\label{rhicsps1}
\end{figure}

It is constructive to compare the pressure of the hadron
gas at freeze-out to the pressure of the ideal gas of quarks and
gluons at the same temperature and baryon chemical potential. To do so
we use the formula
\begin{equation}
P_{QGP}(T,\mu_q)={1 \over 3} \left[
16 {\pi^2 \over 30} T^4 + 6 N_f \left(
{7 \pi^2 \over 120} T^4 + {1\over 4} \mu^2_q T^2 +
{1\over 8 \pi^2} \mu^4_q \right) \right] -B,
\label{pqgp}
\end{equation}
where $N_f$ is the number of quark flavors and $B$ is the bag
constant.  For $T=165$ MeV, $\mu_q=\mu^B/3=41/3$ MeV (the quark chemical
potential is one third of the baryon chemical potential), $B^{1/4}=$
200 MeV, and $N_f=3$ one finds $P=0.3$ GeV/fm$^3$. This result shows
that the pressure of the ideal gas of quarks and gluons is much larger
than the pressure of a hadron gas (we recall that our calculation
gives $P=0.08$ GeV/fm$^3$).  If the equation of state (\ref{pqgp}) was
realistic, we would deal with the plasma rather than with a hadron gas
at the temperatures as high as $T=165$ MeV.  More realistic equations
of state of the plasma include interactions of quarks and gluons which
lower the pressure of the plasma and increase the temperature of the
possible phase transition. The most reliable predictions concerning
the phase transition are obtained by the numerical studies of QCD on a
discretized space-time lattice \cite{Karsch}. The most recent values
of the critical temperature are: $T_c = 173 \pm 8$ MeV for $N_f=2$,
and $T_c = 154 \pm 8$ MeV for $N_f=3$. It is intriguing that the
thermal fits of the particle ratios yield the temperature, which is
very close to the critical temperature inferred from the lattice
simulations. The closeness of these two temperatures suggests that the
chemical content of the hadronic fireball is established just after
hadronization phase transition or, some authors argue, during the
hadronization process itself.  In the latter case the thermal
distributions of hadrons have little in common with
rescattering. Note, however, that our fit is consistent (within
errors) with a temperature smaller than $T_c$.

\begin{figure}[h]
\epsfysize=8cm
\par
\begin{center}
\mbox{\epsfbox{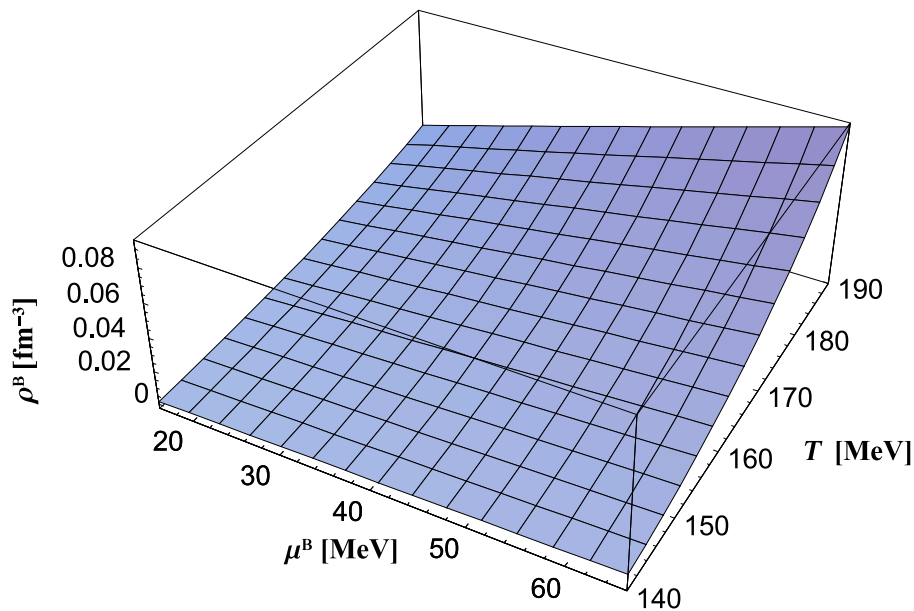}}
\end{center}
\caption{{\small Baryon density of a hadron gas in the vicinity of the
freeze-out point: $T_{chem}=165$ MeV and $\mu^B_{chem}=41$ MeV. }}
\label{rhocont}
\end{figure}

\begin{figure}[h]
\epsfysize=8cm
\par
\begin{center}
\mbox{\epsfbox{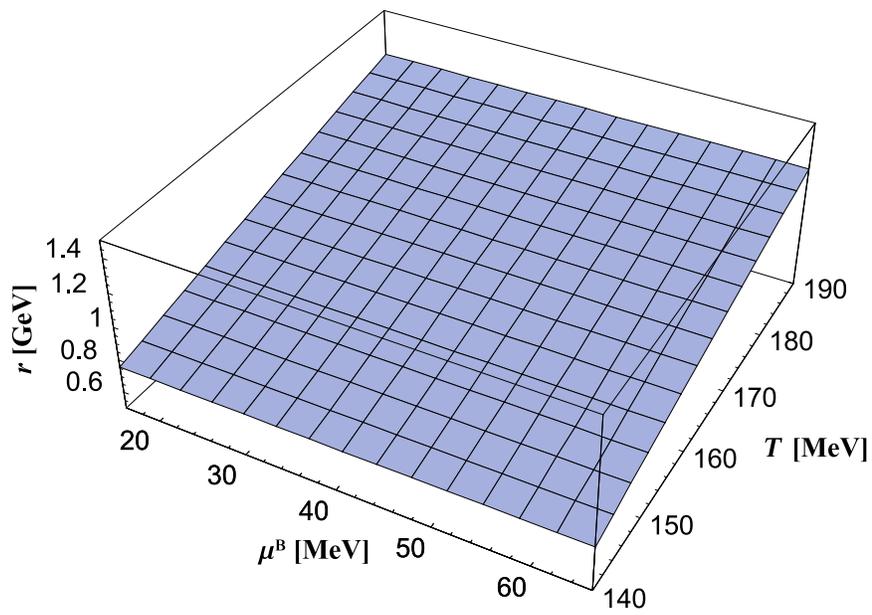}}
\end{center}
\caption{{\small Cleymans-Redlich ratio for a hadron gas. }}
\label{rr}
\end{figure}

We close this Section with the two additional plots.  In
Fig. \ref{rhocont} we show the baryon density of a hadron gas in the
region close to our optimal thermodynamic parameters. In the
considered region the baryon density is very low. It is much smaller
than the baryon saturation density $\rho_0=0.17$ fm$^{-3}$ (we note
that for Pb + Pb collisions at CERN SPS the fitted baryon density was
close to $\rho_0$). Finally, in Fig. \ref{rr} we plot the
Cleymans-Redlich ratio. One can observe that $r$ is close to unity ($
0.8 < r < 1.2 $) in the rather vast range of the temperature and
the baryon chemical potential.

\chapter{Conclusions}

\vspace{0.5cm}

In the end of the paper we summarize the main findings of our work:

\vspace{0.5cm}

\begin{itemize}

\item Thermal description of the particle ratios, based on the
assumption of {\it the full chemical and thermal equilibrium}, works
well for Pb + Pb collisions at CERN SPS and Au + Au collisions at BNL
RHIC. At smaller beam energies and for smaller sizes of the colliding
systems this version of the thermal model reproduces only a
qualitative behavior of the data. We stress that this observation
follows from the study which uses exactly the same framework for all
reactions.

\item The optimal values of the temperature inferred from the thermal
analysis of the ratios in Pb + Pb collisions at CERN SPS and Au + Au
collisions at BNL RHIC agree. The more energetic collisions at RHIC do
not produce a hotter hadronic system. The fitted value of the
temperature ($T_{chem} \sim 165$ MeV for RHIC and $T_{chem} \sim 168$
MeV for SPS), is very close to the critical temperature obtained from
the lattice simulations of QCD.

\item In the case of Pb + Pb collisions at CERN SPS energies, the
correct reconstruction of the weak decays is essential for the
estimates of the energy density and the baryon number density
characterizing the freeze-out. On the other hand, at RHIC the role of
the weak decays is less important.

\item Our calculations confirm the Cleymans-Redlich observation that
thermal conditions at freeze-out, for different colliding systems at
different energies, correspond to an average hadron energy 1 GeV.
Moreover, we have found that averaged baryon energy is 1.6 GeV and the
average meson energy is 0.9 GeV for both Pb + Pb collisions at CERN
SPS and Au + Au collisions at BNL RHIC. This new observation is parallel
to the argument that two different Hagedorn temperatures are required
to describe the mass spectra of baryons and mesons.

\item Possible modifications of the hadron masses and widths were
incorporated into a thermal analysis of the particle ratios. In the
case of Pb + Pb collisions at CERN SPS, we have found that moderate,
up to 20\%, dropping  of the masses does not spoil the quality of the fits.
Larger dropping or growing  of the masses are not likely, since they result in
a significant increase of the $\chi^2$ values. We have also showed
that the increase of hadron widths by less than a factor of two does
not affect the thermal-model fits. Similar behavior is found for
Au + Au collisions at BNL RHIC. In this case, however, larger dropping
of the masses is allowed.

\item We have checked, for a variety of colliding systems, that the
quantum statistics (Bose-Einstein and Fermi-Dirac) can be very well
approximated by the classical Boltzmann statistics. This means that
the hadron gas at the freeze-out behaves like a classical gas with the
equation of state $P = n T$. In addition, the use of the Boltzmann
distribution function implies that the excluded-volume corrections do
not change the values of the fitted optimal values of $T_{chem}$ and
$\mu^B_{chem}$ (provided the eigenvolumes of baryons and mesons are
equal).

\item Finally, we emphasize that our fits may be treated as a first
step in more complex investigations of hadron production in
ultra-relativistic heavy-ion collisions. A combination of the thermal
approach with the hydrodynamic expansion can be used to study other
observables such as: transverse-momentum spectra, rapidity
distributions, elliptic flow, or HBT correlation radii.

\end{itemize}

\newpage

\renewcommand\bibname{Bibliography \\
\bigskip
{\small \hspace{0.75cm} {\underline {\it High-Energy Nuclear Collisions}}}
\vspace{-0.5cm} }

\newpage
\thispagestyle{empty}
$\mbox{}$

\newpage
\thispagestyle{empty}

\vspace{2cm}
\centerline{\bf ACKNOWLEDGMENTS}
\vspace{2cm}

I would like to thank Dr. Wojciech Florkowski for suggesting this
investigation, stimulating remarks, and numerous discussions concerning
the problems presented in this dissertation.

I am also grateful to Dr. Wojciech Broniowski for his interest in this
study, critical comments, and encouragement.

I would like to thank Prof. Edward Kapu\'scik for his help and
guidance during my PhD studies, and Prof. Jan Kwieci\'nski for his
interest in this work. I thank also my colleagues from the
Theory Department of the Institute of Nuclear Physics for interesting
discussions.

Finally, I would like to thank my family, especially my wife Kinga, 
for their understanding, patience and support.


\begin{thebibliography}{999}


\bibitem{QM90}  QUARK MATTER '90. Proceedings, 8th International Conference
on Ultra-Relativistic Nucleus-Nucleus Collisions, MENTON, FRANCE, Nucl.
Phys. {\bf A525} (1991)

\bibitem{QM91}  QUARK MATTER '91. Proceedings, 9th International Conference
on Ultra-Relativistic Nucleus-Nucleus Collisions, GATLINBURG, USA, Nucl.
Phys. {\bf A544} (1992)

\bibitem{QM93}  QUARK MATTER '93. Proceedings, 10th International Conference
on Ultra-Relativistic Nucleus-Nucleus Collisions, BORL\"ANGE, SWEDEN, Nucl.
Phys. {\bf A566} (1994)

\bibitem{QM95}  QUARK MATTER '95. Proceedings, 11th International Conference
on Ultra-Relativistic Nucleus-Nucleus Collisions, MONTEREY, USA, Nucl. Phys.
{\bf A590} (1995)

\bibitem{QM96}  QUARK MATTER '96. Proceedings, 12th International Conference
on Ultra-Relativistic Nucleus-Nucleus Collisions, HEIDELBERG, GERMANY, Nucl.
Phys. {\bf A610} (1996)

\bibitem{QM97}  QUARK MATTER '97. Proceedings, 13th International Conference
on Ultra-Relativistic Nucleus-Nucleus Collisions, TSUKUBA, JAPAN, Nucl.
Phys. {\bf A638} (1998)

\bibitem{QM99}  QUARK MATTER '99. Proceedings, 14th International Conference
on Ultra-Relativistic Nucleus-Nucleus Collisions, TORINO, ITALY, Nucl. Phys.
{\bf A661} (1999)

\bibitem{QM01}  QUARK MATTER '01. Proceedings, 15th International Conference
on Ultra-Relativistic Nucleus-Nucleus Collisions, Brookhaven National 
Laboratory, USA, Nucl. Phys. {\bf A} in print


\vspace{0.5cm} \noindent {\underline{{\em Thermal Models}}} %
\vspace{0.5cm}

\bibitem{PBM-AGS}  P. Braun-Munzinger, J. Stachel, J. P. Wessels, and N. Xu, Phys. Lett.
{\bf B344} (1995) 43


\bibitem{CleyEllSa}  J. Cleymans, D. Elliott, H. Satz, and R. L. Thews, Z. Phys. {\bf C74}
(1997) 319


\bibitem{PBM-SPS}  P. Braun-Munzinger, J. Stachel, J. P. Wessels, and N. Xu, Phys. Lett.
{\bf B365} (1996) 1

\bibitem{RafLetTou}  J. Rafelski, J. Letessier, and A. Tounsi, Acta Phys. Pol. {\bf B28}
(1997) 2841

\bibitem{PBMHS}  P. Braun-Munzinger, I. Heppe, and J. Stachel, Phys. Lett. {\bf B465}
(1999) 15

\bibitem{CleyRed}  J. Cleymans and K. Redlich, Phys. Rev. Lett. {\bf 81} (1998) 5284

\bibitem{YenGor}  G. D. Yen and M. I. Gorenstein, Phys. Rev. {\bf C59} (1999) 2788

\bibitem{BecCleyKSR1}  F. Becattini, J. Cleymans, A. Keranen, E. Suhonen, and K. Redlich, Phys. Rev.
{\bf C64} (2001) 024901

\bibitem{BecCleyKSR2}  F. Becattini, J. Cleymans, A. Keranen, E. Suhonen, and K. Redlich,
hep-ph/0011322

\bibitem{Gaz}  M. Ga\'zdzicki, Nucl. Phys. {\bf A681} (2001) 153

\bibitem{nxu}  N. Xu and M. Kaneta, nucl-ex/0104021, Proceedings of QM2001, Nucl. Phys. A in print

\bibitem{pbmrhic}  P. Braun-Munzinger, D. Magestro, K. Redlich, and J. Stachel, Phys. Lett.
{\bf B518} (2001) 41

\bibitem{FlorBronMich-TherAnal}  W. Florkowski, W. Broniowski, and M. Michalec, 
nucl-th/0106009

\bibitem{sh} J. Rafelski and J. Letessier, Phys. Rev. Lett. {\bf 85} (2000) 4695

\bibitem{ptspec} W. Broniowski and W. Florkowski, nucl-th/0106050, Phys. Rev. Lett.
in print


\vspace{0.5cm} \noindent {\underline{{\em Models of QGP Production}}} %
\vspace{0.5cm}

\bibitem{tubes} A. Bialas, W. Czy\.z, A. Dyrek, and W. Florkowski,
Nucl. Phys. {\bf B296} (1988) 611; A. Dyrek and W. Florkowski, Nuovo
Cim. {\bf 102A} (1989) 1013

\bibitem{cascade} K. Geiger, Phys. Rep. {\bf 258} (1995) 237


\vspace{0.5cm} \noindent {\underline{{\em Hadrons in Medium}}} %
\vspace{0.5cm}

\bibitem{Hirsch95}  {\it Hadrons in Nuclear Matter}, Proceedings of the International
Workshop XXIII on Gross Properties of Nuclei and Nuclear Excitations, Hirschegg, AUSTRIA, 1995

\bibitem{Hirsch00}  {\it Hadrons in Dense Matter}, Proceedings of the International
Workshop XXVIII on Gross Properties of Nuclei and Nuclear Excitations, Hirschegg, AUSTRIA, 2000

\bibitem{hatsuda}  T. Hatsuda, H. Shiomi, and H. Kuwabara, Prog. Theor. Phys. {\bf 95} (1996) 1009

\bibitem{klingl}  F. Klingl, N. Kaiser, and W. Weise, Nucl. Phys. {\bf A624} (1997) 527

\bibitem{HatsudaLee}  T. Hatsuda and S. H. Lee, Phys. Rev. {\bf C46} (1992) R34

\bibitem{BR}  G. Brown and M. Rho, Phys. Rev. Lett. {\bf 66} (1991) 2720

\vspace{0.5cm} \noindent {\underline{{\em Relativistic Hydrodynamics}}} %
\vspace{0.5cm}

\bibitem{Fermi}  E. Fermi, Prog. Theor. Phys. {\bf 5} (1950) 570; Phys. Rev.
{\bf 92} (1953) 452

\bibitem{Landau}  L.D. Landau, Izv. Akad. Nauk SSSR, Ser. Fiz., {\bf 17}
(1953) 51, also published in ``Collected papers of L.D. Landau'', ed. by D.
Ter Haar, Pergamon Press (1965) 569

\bibitem{Bjorken}  J. Bjorken, Phys. Rev. {\bf D27} (1983) 140

\bibitem{Baym}  G. Baym, B. Friman, J.-P. Blaizot, M. Soyeur and W. Czy\.z,
Nucl. Phys. {\bf A407} (1983) 541

\bibitem{Heinz}  U. Heinz, J. Phys. {\bf G25} (1999) 263

\bibitem{CleyOesRed}  J. Cleymans, H. Oeschler, and K. Redlich, J. Phys. {\bf G25} (1999) 281

\bibitem{CooperFrye}  F. Cooper, G. Frye, Phys. Rev. {\bf D10} (1974) 186
\bibitem{CooperFryeSchon}  F. Cooper, G. Frye, and E. Schonberg, Phys. Rev. {\bf D11} (1975) 192


\vspace{0.5cm} \noindent {\underline{{\em Particle Data Tables}}} %
\vspace{0.5cm}

\bibitem{PDG} Particle Data Group, Eur. Phys. J. {\bf C15} (2000) 1


\vspace{0.5cm} \noindent {\underline{{\em Modifications of Hadron Properties at Freeze-Out}}} %
\vspace{0.5cm}

\bibitem{zsch}  D. Zschiesche, L. Gerland, S. Schramm, J. Schaffner-Bielich, H. Stoecker,
and W. Greiner, Nucl. Phys. {\bf A681} (2001) 34

\bibitem{MichFlorBron-Scal}  M. Michalec, W. Florkowski, and W. Broniowski, 
Phys. Lett. {\bf B520} (2001) 213

\bibitem{FlorBron}  W. Florkowski, W. Broniowski, Phys. Lett. {\bf B477} (2000) 73

\bibitem{Hirsch}  W. Florkowski and W. Broniowski, Proceedings of the International Workshop
XXVIII on Gross Properties of Nuclei and Nuclear Excitations, Hirschegg, AUSTRIA, 2000, p. 275




\vspace{0.5cm} \noindent {\underline{{\em Hagedorn Temperature}}} %
\vspace{0.5cm}

\bibitem{hag2}  W. Broniowski and W. Florkowski, Phys. Lett. {\bf B490} (2000) 223

\bibitem{brohag1}  W. Broniowski, {\it Few-Quark Problems}, Proceedings of the Mini-Workshop,
Bled, Slovenia, 2000, p. 14

\bibitem{brohag2}  W. Broniowski, Acta Phys. Pol. {\bf B31} (2000) 2155

\bibitem{wb} W. Broniowski, private communication

\vspace{0.5cm} \noindent {\underline{{\em Lattice QCD}}} %
\vspace{0.5cm}

\bibitem{Karsch}  F. Karsch, hep-ph/0103314, Proceedings of QM2001, Nucl. Phys. {\bf A} in print


\vspace{0.5cm} \noindent {\underline{{\em Finite-Size and Excluded-Volume Corrections}}} %
\vspace{0.5cm}

\bibitem{JMZ}  H. R. Jaqama, A. Z. Mekijan, and L. Zamick, Phys. Rev. {\bf C29} (1984) 2067,
see also R. Balian and C. Bloch, Ann. Phys. {\bf 70} (1970) 401

\bibitem{YGGY}  G. D. Yen, M. I. Gorenstein, W. Greiner, S.-N. Yang, Phys. Rev. {\bf C56} (1997)
2210

\vspace{0.5cm} \noindent {\underline{{\em CERN Press Release Feb. 10, 2000}}} %
\vspace{0.5cm}

\bibitem{press}  http://cern.web.cern.ch/CERN/Announcements/2000/NewStateMatter/

\vspace{0.5cm} \noindent {\underline{{\em Experiments on Dilepton Production}}} %
\vspace{0.5cm}

\bibitem{CERES} CERES Collab., G. Agakichiev {\it et al.}, Phys. Rev. Lett. {\bf 75} (1995) 1272

\bibitem{HELIOS}  HELIOS/3 Collab., M. Masera {\it et al.}, Nucl. Phys. {\bf A590} (1995) 93c

\newpage

\vspace{0.5cm} \noindent {\underline{{\em Chiral Models at Finite Temperature/Density }}} %
\vspace{0.5cm}

\bibitem{c1} M. Lutz, S. Klimt, and W. Weise, Nucl. Phys. {\bf A542} (1992) 521; M. Lutz, A.
Steiner, and W. Weise, Nucl. Phys. {\bf A574} (1994) 755

\bibitem{c2} T. Hatsuda and T. Kunihiro, Phys. Rep. {\bf 247} (1994) 221, and references 
therein


\vspace{0.5cm} \noindent {\underline{{\em Thermodynamics of Hadron Gas}}} %
\vspace{0.5cm}

\bibitem{BU}  E. Beth and G. E. Uhlenbeck, Physica {\bf 3} (1936) 729; Physica {\bf 4} (1937) 915

\bibitem{DMB}  R. Dashen, S.-K. Ma and H. J. Bernstein, Phys. Rev. {\bf 187} (1969) 345

\bibitem{WW}  W. Weinhold, {\it Zur Thermodynamik des $\pi$N-Systems}, Diplomarbeit,
GSI, Sept. 1995

\bibitem{WFN}  W. Weinhold, B. Friman, and W. Norenberg, Acta Phys. Pol. {\bf B27} (1996) 3249;
Phys. Lett. {\bf B433} (1998) 236

\vspace{0.5cm} \noindent {\underline{{\em Experimental Results for RHIC}}} %
\vspace{0.5cm}

\bibitem{phobos}  B. B. Back {\it et al.}, PHOBOS Collaboration, Phys. Rev. Lett. {\bf 85} 
(2000) 3100

\bibitem{bearden}  I. G. Bearden, BRAHMS Collaboration, Proceedings of QM2001, Nucl. Phys.
{\bf A} in print

\bibitem{harris}  J. Harris, STAR Collaboration, Proceedings of QM2001, Nucl. Phys. {\bf A} in print

\bibitem{caines}  H. Caines, STAR Collaboration, Proceedings of QM2001, Nucl. Phys. {\bf A} in print

\bibitem{ohnishi} H. Ohnishi, PHENIX Collaboration, Proceedings of QM2001, Nucl. Phys.
{\bf A} in print

\bibitem{zxu} Z. Xu, nucl-ex/0104001, Proceedings of QM2001, Nucl. Phys. {\bf A} in print



\end{thebibliography}
\end{document}